\documentclass[11pt,final]{article}
\pdfoutput=1
\usepackage{MRnotes}

\newcommand{\RNAdS}[1]{Reissner-Nordstr\"om-AdS$_{#1}$}

\newcommand{\RQ}{r_{_{\!Q}}}    
\newcommand{\bRQ}{\mathfrak{C}}   
\newcommand{\sdc}{\mathfrak{S}_{_Q}}  
\newcommand{\Fint}[1]{\mathfrak{T}\left[#1\right]}  
\newcommand{\FHint}[1]{\mathfrak{H}\left[#1\right]}  

\newcommand{\VV}{\mathbb{V}}  
\newcommand{\ScS}{\mathbb{S}} 
\newcommand{\ai}{\alpha}   
\newcommand{\VVV}{\mathbb{X}}
\newcommand{\VVSa}{\mathbb{A}}
\newcommand{\VVSb}{\widetilde{\mathbb{A}}}
\newcommand{\VSSa}{\mathbb{B}}
\newcommand{\VSSb}{\widetilde{\mathbb{B}}}

\newcommand{\Mser}[2]{\varphi_{_{#1}}^{#2}}  
\newcommand{\Dfn}[2]{\Delta_{_{#1}}^{#2}} 
\newcommand{\Dfnh}[2]{\hat{\Delta}_{_{#1}}^{#2}} 
\newcommand{\xser}[1]{\varphi_{_{\MX,\eta}}^{#1}}  
\newcommand{\yser}[1]{\varphi_{_{\MY,\eta}}^{#1}}  
\newcommand{\yserh}[1]{\hat{\varphi}_{_{\MY,\eta}}^{#1}}  

\newcommand{\ann}{\mathscr{M}}   

\newcommand{\bwt}{\mathfrak{w}}   
\newcommand{\bqt}{\mathfrak{q}}    
\newcommand{\bk}{\vb{k}}  
\newcommand{\bx}{\vb{x}}   
\newcommand{\sk}{\mathsf{k}} 

\newcommand{\ctor}{\zeta}  
\newcommand{\Dz}{\mathbb{D}}  

\newcommand{\In}{\text{\tiny{in}}}     
\newcommand{\Out}{\text{\tiny{out}}}  

\newcommand{\nB}{n_{_B}}   
\newcommand{\Gin}[1]{G_{_{#1}}^\In}     
\newcommand{\Kin}[1]{K_{_{#1}}}      
\newcommand{\Gout}[1]{G_{_{#1}}^\Out}   
\newcommand{\JF}{J_{_{\text{F}}}}
\newcommand{\JP}{J_{_{\text{P}}}}

\newcommand{\MX}{\mathscr{X}}   
\newcommand{\MY}{\mathscr{Y}}   

\newcommand{\MV}{\mathscr{V}}
\newcommand{\MZ}{\mathscr{Z}}
\newcommand{\Vd}{\mathsf{V}}
\newcommand{\Zd}{\mathsf{Z}} 
\newcommand{\MW}{\Theta}
\newcommand{\VZd}{\mathfrak{V}_{\mathsf{Z}}}
\newcommand{\VVd}{\mathfrak{V}_{\mathsf{V}}}
\newcommand{\Lk}{\Lambda_k}

\newcommand{\Np}[1]{\mathsf{N}_{_#1}}  

\newcommand{\BQT}{\mathfrak{p}_s}    
\newcommand{\BQTCS}{\mathfrak{p}_{\eta}}    
\newcommand{\kcs}{\kappa_{_\text{CS}}}

\newcommand{\POp}{\mathcal{P}}  
\newcommand{\QOp}{\mathcal{Q}}  
\newcommand{\EOp}{\mathcal{E}}  

\newcommand{\EEin}{\mathcal{E}^{\text{\tiny{Ein}}}}   
\newcommand{\EMax}{\mathcal{E}^{\text{\tiny{Max}}}}  

\newcommand{\vGR}{\Psi}   
  
\newcommand{\vMax}{\Xi}     

\newcommand{\PHW}{\Phi_{_\text{W}}}

\newcommand{\skR}{\text{\tiny R}}  
\newcommand{\skL}{\text{\tiny L}}  

\title{Anomalous hydrodynamics effective actions from holography}
\author[a]{Mukund Rangamani,}
\author[a]{Julio Virrueta,}
\author[b]{Shuyan Zhou}

\affiliation[a]{
	Center for Quantum Mathematics and Physics (QMAP)\\
	Department of Physics \& Astronomy, University of California, Davis, CA 95616 USA.}
\affiliation[b]{
		 Department of Physics, Fudan University, Shanghai 200433, China.}
\emailAdd{mukund@physics.ucdavis.edu}
\emailAdd{jvirrueta@ucdavis.edu}
\emailAdd{shuyanzhou20@fudan.edu.cn}
\abstract{
We derive an effective action for charged plasmas with an anomalous (abelian) global current charge current using holography. The holographic description is captured by the dynamics of an Einstein-Maxwell-Chern-Simons theory in an asymptotically AdS spacetime. The 't Hooft anomaly contribution, which is  encoded in the Chern-Simons term, contributes at the Gaussian order in the effective action only in the momentum diffusion sector, where it leads to chiral shear waves. However, as we demonstrate, beyond the Gaussian order, there is non-trivial imprint of the parity-violating anomaly term in sound and charge diffusion dynamics as well. 
}

\begin{document}
\maketitle
 

\section{Introduction}\label{sec:intro}

The fact that quantum anomalies can have non-trivial imprint on hydrodynamic transport, first made manifest in the holographic computation of~\cite{Banerjee:2008th,Erdmenger:2008rm}, is an important lesson arising from the fluid/gravity correspondence~\cite{Bhattacharyya:2008jc}. We now understand, thanks to the elegant argument of~\cite{Son:2009tf}, that positivity of local entropy production implies that the quantum anomaly fixes parity violating transport data in the fluid. Subsequent developments have resulted in a comprehensive physical picture for anomalous hydrodynamics from various perspectives, e.g., at the level of constitutive relations~\cite{Loganayagam:2011mu}, using Kubo formulae~\cite{Amado:2011zx,Landsteiner:2012kd}, or directly from a Euclidean partition function~\cite{Banerjee:2012iz,Banerjee:2012cr,Jensen:2012kj,Bhattacharyya:2013ida}, etc. A useful perspective on the problem is to use the inflow mechanism, which coupled with Euclidean consistency arguments, allows for a complete solution to anomaly induced transport ~\cite{Jensen:2013kka,Jensen:2013rga}.  A review of these developments can be found in~\cite{Landsteiner:2016led}.

Of present interest is the simplest incarnation of these effects, viz., the 't Hooft anomaly of an abelian global symmetry in four spacetime dimensions. As a concrete example, consider the $\mathrm{SU}(N)$ $\mathcal{N}=4$ Super Yang-Mills theory with its anomalous $\mathrm{SU}(4)_R$ symmetry. We will focus on a $\mathrm{U}(1)_R$ subgroup for simplicity. This theory in the large $N$, strong coupling limit is holographic to Einstein-Maxwell-Chern-Simons theory, with the Chern-Simons term capturing the 't Hooft anomaly through an AdS/CFT version of the inflow mechanism. Indeed, it was the analysis of this theory in the fluid/gravity context~\cite{Banerjee:2008th,Erdmenger:2008rm}, which led to the progress on understanding anomalous transport.

A natural question is whether one can alternatively incorporate such effects into a hydrodynamic effective action. This problem was successfully addressed\footnote{ 
 	The first attempt to construct anomalous hydrodynamic effective actions in~\cite{Dubovsky:2011sk} captured the physics of abelian anomalies in two dimensions, but encountered obstructions more generally.} 
in~\cite{Haehl:2013hoa} who obtained the anomalous part of the hydrodynamic action directly from the anomaly polynomial. In particular, their action was given as the difference of two Chern-Simons forms, i.e., in terms of a transgression form.\footnote{
 	A transgression form can be defined in terms of an interpolating connection between two gauge potentials~\cite{Bertlmann:2000ano}. It is equal to the difference of two Chern-Simons forms up to an exact form. } 
Curiously, while the transgression form captured the constitutive relations accurately,  in order to obtain the correct anomalous conservation equations,~\cite{Haehl:2013hoa} argued in favour of a Schwinger-Keldysh effective action with a non-trivial influence functional.

An effective action for \emph{dissipative hydrodynamics} should necessarily involve non-trivial influence functions to ensure that the fluctuation-dissipation relations are upheld. However, is surprising at first sight why this must be the case for the non-dissipative anomalous transport. This seems especially puzzling, since the anomaly induced transport can be directly recovered from a Euclidean partition function. Nevertheless, there is a sense in which the anomaly induced transport serves as a prototype for more general phenomena in hydrodynamics. As argued in~\cite{Haehl:2015pja}, the adiabatic (i.e., non entropy producing) part of hydrodynamics includes many classes  of transport which demand an effective action comprising non-trivial influence functionals (and the introduction of a thermal gauge symmetry). Our current understanding of hydrodynamic effective field theories initiated in~\cite{Crossley:2015evo,Haehl:2015uoc} captures these features as demonstrated in~\cite{Jensen:2018hse}.

From a physical perspective it would be helpful to understand the rationale for why the anomaly contributions to the effective action necessitates a non-trivial influence functional. We take a step in this direction by leveraging holography to construct an anomalous hydrodynamic effective action. While the aforementioned earlier literature constructs a non-linear effective action, whose variation with respect to background sources directly gives the constitutive relations, our focus here will be on constructing an action directly in terms of the physical moduli, perturbatively in an amplitude expansion. It should, however, be noted that the derivation we present herein is a-priori logically distinct from the earlier discussion of~\cite{Haehl:2013hoa}, and importantly provides a independent cross-check of the earlier analysis.

Non-linear hydrodynamic effective actions focus on constructing a dynamical system whose equations of motion directly result in the conservation equations for the currents. This is achieved using a sigma model approach, where the hydrodynamic fields are maps from a worldvolume to the physical fluid. Linearized solutions of these conservation equations (say in flat spacetime) include momentum diffusion (shear), energy transport (sound), and charge diffusion modes. The shear modes are transverse to direction of propagation, while the sound and charge diffusion are longitudinal. One can parameterize the currents and the effective action directly in terms of these modes treating them as the low-lying moduli fields,  cf.,~\cite{Ghosh:2020lel,He:2022jnc,He:2021jna,He:2022deg}.  The non-linear nature of hydrodynamics is manifested by the resulting action having coupling of arbitrary order between these fields. We construct the effective action for a four-dimensional relativistic fluid with an anomalous $U(1)$ global symmetry, capturing all the leading vertices that depend on the 't Hooft anomaly coefficient.\footnote{
	In our analysis we will not be turning on sources for the hydrodynamic variables, choosing instead to work directly with the physical fluctuation modes, as noted. This in particular implies that for the most part we will be insensitive to the difference between the consistent and covariant currents (see also~\cite{Banerjee:2008th}).}  

The structural part of our result is universal and applies to any relativistic charged fluid. We will present the result directly in terms of the transport data, confirming the same with explicit values using a particular holographic embedding. A 4d relativistic fluid has five low-lying moduli: 2 momentum diffusion modes $\POp_\eta$, a single charge diffusion mode $\QOp$, and a sound mode $\EOp$ (which counts as two degrees of freedom). As in~\cite{He:2021jna,He:2022deg}, we  parameterize the conserved current in terms of these fields. The remaining components of the currents are non-hydrodynamic -- they decay on short timescales. The 't Hooft anomaly of the global current causes the momentum diffusion modes to be chiral, $\POp_\pm$, but leaves the other modes unaffected~\cite{Sahoo:2009yq} (see also~\cite{Bhattacharyya:2013ida}). 

The real-time Schwinger-Keldysh effective action for the system takes the following schematic form in the average/difference basis:\footnote{
	The $4$-momenta are denoted by $\sk^\mu = (\omega,\vb{k})$, and $k=\abs{\vb{k}}$ is the spatial momentum magnitude. We often suppress the Lorentz index, and write expressions using dimensionless frequencies and momenta defined in~\eqref{eq:bwqdef}.
	When necessary, frequency reversed $4$-momentum is singled out by a bar decoration, i.e., $\bar{\sk}^\mu= (-\omega,\vb{k})$. Finally, the Fourier space integrals are indicated by a shorthand notation, 
	$\int_\sk \equiv \int \frac{d\omega}{2\pi}\, \int \frac{d^3\vb{k}}{{(2\pi)}^3}$.
}
\begin{equation}\label{eq:SanomA}
\begin{split}
S_{_\text{WIF}}\left[\POp_\eta,\QOp,\EOp\right] 
&= 
	\sum_{\mathcal{O}\in \{\POp_\eta,\QOp,\EOp\}} \int_\sk\,(- k^2 )\left[  \mathcal{O}_d^\dagger \, \Kin{\mathcal{O}}\left(\mathcal{O}_a + \left(\nB+\frac{1}{2}\right)\mathcal{O}_d   \right) + \text{cc} \right] +S_\text{int}  . 
\end{split}
\end{equation}	
The kinetic operators $\Kin{\POp_\eta}$, $\Kin{\QOp}$ and $\Kin{\EOp}$ are the chiral shear, charge diffusion, and sound dispersion relations, In fact, as argued above,  the latter two dispersion functions are unchanged from the results obtained in~\cite{He:2022deg}.  The new result is  the determination of the chiral shear dispersion to quartic order in gradients, extending the result of~\cite{Sahoo:2009yq}.  Additionally, we explore the qualitative features of the non-Gaussian terms contained in $S_\text{int}$.  We give a general expression for  $S^{(3)}_{_\text{WIF,anom}}$, the cubic contribution that is linear in the anomaly coefficient, and compute it explicitly at the first few orders in the  hydrodynamic gradient expansion. The results we derive herein reproduce the known transport coefficients constrained by the 't Hooft anomaly.
However, we emphasize that these results are new, and have not hitherto appeared in the literature.

The hydrodynamic effective action at cubic order has additional parity-even contributions. These exist for any charged plasma anomalous or otherwise. As our focus is on the anomaly constrained part of the action, we do not explore such terms in the current work.  Furthermore, we also note that there are higher order non-Gaussian terms, which are non-linear in the anomaly coefficient. These, however, are not solely determined by the anomaly. While they are anomaly induced, their origins lie in the fact that they are admissible owing to the parity-breaking in the system, but are not completely constrained by the anomaly. They can be changed for instance by other higher derivative bulk interactions.  While the primary focus here is to derive~\eqref{eq:SanomA} holographically, we do compare the result with the earlier study of the anomalous hydrodynamic effective action~\cite{Haehl:2014zda} (further aspects of the connection to hydrodynamic effective theories will be developed in~\cite{Loganayagam:2023xyz}).

The outline of the paper is as follows: In~\cref{sec:setup} we outline the basic holographic set-up of the problem. Parameterizing the linearized fluctuations in terms of the physical hydrodynamic modes, we derive the quadratic and cubic parts of the bulk actions. Using these results, in~\cref{sec:anomalySeff} we obtain the Gaussian effective action and  the leading anomaly-induced non-Gaussian corrections thereto. Finally, in~\cref{sec:discuss} we end with a discussion of the results. Since the major part of our analysis parallels that of~\cite{He:2021jna,He:2022deg} we have chosen to be succinct in our description of the technicalities. We have, nevertheless, collated some relevant information in~\cref{sec:vectors}. Explicit solutions to the shear mode dynamics can be found in~\cref{sec:vecXYsols}. Finally,~\cref{sec:harmonics} includes some useful data of chiral vector harmonics in 4 spacetime dimensions, and~\cref{sec:mockT} gives the expression for the mock tortoise coordinate in \RNAdS{5} (which is relevant for the computation of non-Gaussian terms).

\section{The gravitational model}\label{sec:setup}

Consider the Einstein-Maxwell-Chern-Simons (EMCS) theory, with action 
\begin{equation}\label{eq:SEMax}
\begin{split}
S_\text{EM} 
&= 
	\frac{1}{16\pi G_N}\, \int d^5 x\, \sqrt{-g} \, 
		\left[ R + 12- \frac{1}{2} \,F_{AB}\, F^{AB} 
        + \frac{2}{3}\, \kcs\, \epsilon^{ABCDE}\, A_A\, F_{BC}\, F_{DE}
        \right] \\
&\qquad  \qquad    
    + S_\text{bdy} + S_\text{ct} \,,\\
S_\text{bdy}
&=  \frac{1}{8\pi G_N}\, \int d^4 x\, \sqrt{-\gamma} \, K		  \,.
\end{split}
\end{equation}
Here $g_{AB}$ is the metric, $\gamma_{\mu\nu}$ the induced metric on the timelike asymptotic boundary, and $K$ is the extrinsic curvature of the boundary.\footnote{ 
	We employ the same conventions as~\cite{He:2021jna} for ease of comparison. Specifically, the  AdS length scale is set to unity $\lads =1$. Index conventions: uppercase Latin alphabet ($A,B,\cdots$) to indicate bulk spacetime indices,  Greek alphabets ($\mu,\nu, \cdots$) will refer to boundary spacetime indices, while lowercase Latin alphabets ($i,j, \cdots$) will be used to refer to the spatial directions along the boundary.
 \label{fn:conventions}}  The counterterm action $S_\text{ct}$ is necessary to obtain finite physical answers for boundary observables. The equations of motion  from~\eqref{eq:SEMax} are 
\begin{equation}\label{eq:EMeqns}
\begin{split}
\EEin_{AB}&\equiv
R_{AB} - \frac{1}{2}\, R\, g_{AB} - 6\, g_{AB} 
=
	 g^{CD}\, F_{AC} \, F_{BD} - \frac{1}{4}\, g_{AB}\, F_{CD}\, F^{CD} \,,\\
\EMax_B&\equiv
\nabla_B F^{AB} - \kcs \, \epsilon^{ABCDE}\, F_{BC}\, F_{DE} = 0	\,.
\end{split}
\end{equation}
This classical system is a consistent truncation of Type IIB supergravity; it  captures the dynamics of the stress tensor and an anomalous $U(1)_R$ current of 4d $\mathcal{N}=4$ SYM.\footnote{
	The field theory central charge is $c_\text{eff} = \frac{\lads^3}{16\pi\,G_N} = \frac{N^2}{8\pi^2}$. The 't Hooft anomaly coefficient is 
	$c_A = \frac{N^2}{12\pi^2}\, \kcs$, with $\kcs = \frac{1}{2\sqrt{6}}$ with our normalization conventions.  We will however leave $\kcs$ arbitrary for much of our discussion to facilitate isolating the anomalous contributions. } 
We will be interested in linearized perturbations around the \RNAdS{5} spacetime, which is a particular solution of~\eqref{eq:EMeqns}, specified by 
\begin{equation}\label{eq:RNAdS}
\begin{split}
ds_{(0)}^2 
&=
	2 dv dr - r^2\, f(r)\, dv^2 + r^2\, d\vb{x}^2 \,,  \hspace{2cm}
	\vb{A}_{(0)} = -a(r)\, dv \,,\\
f(r) 
&= 
	1 - (1+Q^2)\,\left(\frac{r_+}{r}\right)^4 + Q^2\, \left(\frac{r_+}{r}\right)^{6}\,, \hspace{1cm} 
	a(r) = \sqrt{\frac{3}{2}}\, Q\, \frac{r_+^3}{r^2} \,.
\end{split}
\end{equation}	

This geometry is dual to the Gibbs state of the CFT with 
\begin{equation}\label{eq:TRN}
T =  \frac{2- Q^2}{2\pi}\, r_+  \,, \qquad \mu =\sqrt{\frac{3}{2}} \, Q \, r_+  \,. 
\end{equation}	
This equilibrium state, which we expand around, can equivalently be  characterized by one-point functions of the conserved currents
\begin{equation}\label{eq:idealfluid}
\begin{split}
 T_{\mu\nu} 
&=  
	\frac{N^2}{8\pi^2}\, (1+  Q^2)\, r_+^4  \left(\eta_{\mu\nu} + 4\, u_\mu\, u_\nu\right) \,, \qquad 
J_\mu 
= 
	 \frac{N^2}{4\pi^2}\,\sqrt{6} \, Q \, r_+^{3} \, u_\mu\,,
\end{split}
\end{equation}	
with $u^\mu \,u_\mu =-1$ and $u^\mu = \left(\pdv{v}\right)^\mu$ on the boundary. 

\paragraph{Perturbations \& Gaussian dynamics:} The linearized perturbations can be analyzed in terms of gauge invariant variables~\cite{Kodama:2003kk}. We decompose perturbations in planar harmonics on $\mathbb{R}^{3,1}$. Modes which transform under distinct $SO(2)$ (the little group which leaves the spatial momentum vector $\vb{k}$ fixed) 
representations decouple at quadratic order~\cite{He:2021jna,He:2022deg}. We use dimensionless frequency and momenta, scaling out a factor of $r_+$ (which serves as a proxy for the temperature, cf.,~\eqref{eq:TRN})
\begin{equation}\label{eq:bwqdef}
\bwt = \frac{\omega}{r_+}\,, \qquad \bqt = \frac{k}{r_+}\,.
\end{equation}	

Of the 8 physical degrees of freedom, four are hydrodynamic:\footnote{ The transverse traceless tensor modes correspond to the propagating spin-2 field, and the two transverse vector modes to the spin-1 field.} two chiral fields $\MX_\eta$ arise from the transverse vector representation ($\eta$ labels the chirality), while two fields $\Vd$ and $\Zd$ arise from the (longitudinal) scalar representation. The parameterization of the gauge field in terms of the hydrodynamic modes is 
\begin{equation}\label{eq:Amaxhydro}
\vb{A}_A^{(1)}\, dx^A 
= 
	 \frac{1}{r}\left(dv\, \Dz_+ - dr\, \dv{r}\right) \MV(r,\sk)\,\ScS -  
	 \frac{\bqt^2}{2\,\mu}(1-h)\, \MX_\eta(r,\sk) \, \VV^\eta_i\, dx^i  \,.
\end{equation}
Here $\Dz_+ = r^2 f\, \dv{r} - i\,\omega $, and the field $\MV$ is a functional linear combination of $\Vd$ and $\Zd$, 
\begin{equation}\label{eq:MZVdiagonal}
\begin{split}
\MV 
&=  
	\left(\frac{a}{\bRQ} - r_+\, \frac{\BQT^2+2}{4}\right) \frac{\Vd}{h} +\left(  a + \frac{r_+\, \bRQ}{4} \, \BQT^2\right) \frac{h}{\Lk} \, \Zd \,.
\end{split}
\end{equation}
We introduced a set of functions, $h(r), \Lk(r)$, and parameters $\RQ,\bRQ$, to characterize the perturbations. They are defined as
\begin{equation}\label{eq:hLkrqC}
\begin{split}
h &= 1-\frac{\RQ^2}{r^2}\,, \hspace{3cm}
\Lk = k^2 + \frac{3}{2}\, r^3\, f' \,, \\ 
\RQ &=\sqrt{\frac{3\,Q^2}{2\,(1+Q^2)}}\,r_+\,, \hspace{1.66cm}
\bRQ = 2\,\sqrt{\frac{2}{3}}\, \frac{1+Q^2}{Q} \,. \\
\end{split}
\end{equation}	
In addition, we have introduced the deformed momenta
\begin{equation}\label{eq:BQTdefs}
\BQT^2 = \sqrt{1+\frac{8}{3}\,\frac{\bqt^2}{\bRQ^2}} -1 \,, \qquad 
\BQTCS^2  = \sqrt{\left(1+4\,\eta\,\kcs\, \frac{\bqt}{\bRQ}\right)^2 + 2\,\frac{\bqt^2}{\bRQ^2}} - \left(1+4\,\eta\,\kcs\, \frac{\bqt}{\bRQ}\right) ,
\end{equation}	
which arise as we decouple the dynamics into the hydrodynamic fields and diagonalize the kinetic term. They are non-linear functions of spatial momentum magnitude. We also see here that the diagonalized variables have a non-trivial dependence on the Chern-Simons coefficient, $\kcs$, which determines the anomaly. 

In parameterizing the gauge field we have made a gauge choice to focus on the modes which contribute to the dynamics. Note that the bulk Chern-Simons term is gauge invariant only up to boundary terms. This boundary term,  from a holographic perspective, is an implementation of the inflow mechanism.  For example, gauge transformation $ \bm{A}^{(1)} \to \bm{A}^{(1)} + d \lambda$ will induce a spatial component $\lambda(r)\, \partial_i \ScS$, and modify the Chern-Simons contribution.  We will work with the variables chosen to extract the dynamical degrees of freedom, but it should be borne in mind that our actions are defined up to such a restriction. It might in a sense be more natural to construct an action as in~\cite{Haehl:2013hoa} taking the inflow picture seriously, but we will not do so here. 

The metric perturbations are a bit more involved. We start with the perturbed metric parameterized as in~\eqref{eq:gABpar}. A series of field redefinitions described in~\cref{sec:vectors} describes how to package this in terms of the fields $\{\MX_\eta,\Vd,\Zd\}$. These fields are distinguished in that they have diagonal kinetic terms.

Ignoring the non-hydrodynamic modes, it turns out that EMCS dynamics can be reduced to the following classical action for the fields $\{\MX_\eta,\Vd,\Zd\}$:\footnote{
	The derivation of the quadratic action for the vector modes in the presence of anomaly is described in~\cref{sec:vectors}. The scalar sector being insensitive to the anomaly (there is no $\kcs$ dependence at quadratic order) can be directly read off from~\cite{He:2022deg}, where the dynamics in \RNAdS{d+1} for $d\geq 3$ was obtained. As we only require the specialization to $d=4$ for the current analysis, we simply collate the final result for quick reference.}
\begin{equation}\label{eq:XVZbulk}
\begin{split}
S[\MX^\pm,\Vd,\Zd]
&	- \int dr\, \int_\sk \, \frac{\sqrt{-g}}{r^6}\, k^2 
	\left[\mathcal{L}_{_\MX} + \mathcal{L}_{_\Vd} + \mathcal{L}_{_\Zd}\right] + S_3[\MX,\Vd,\Zd]	+ S_\text{bdy}[\MX, \Vd,\Zd] \,,\\
\mathcal{L}_{_\MX}
&=
	 \Np{\MX,\eta}(\BQTCS)\, \left(\nabla_A \MX_\eta\, \nabla^A\MX_\eta - \frac{r_+^2\,(1-h)}{r^2}\, \bRQ^2\, \BQTCS^2\,  \MX_\eta \, \MX_\eta \right), \\
\mathcal{L}_{_\Vd}
&=	\Np{\Vd}(\BQT)\,\frac{r^2}{h^2}\,
		 	\left(\nabla_A\Vd\,\nabla^A\Vd -\VVd(\BQT) \, \Vd^2\right) , \\
\mathcal{L}_{_\Zd} 
&= 
	 \frac{1}{6}\, \Np{\Zd}(\BQT)\, \frac{r^2\,h^2}{\Lk^2}
	 \left(\nabla_A\Zd\,\nabla^A\Zd  -\frac{2+\BQT^2}{1+\BQT^2}\,\frac{rf'}{h\,\Lk} \,k^2\, \Zd^2+  \VZd(\BQT)\, \Zd^2 \right) .	
\end{split}
\end{equation}

The potentials $\VVd, \VZd$, while explicit, are messy -- they were obtained in~\cite{He:2022deg} and collated in~\cref{sec:vectors} for quick reference. The normalization factors $\Np{\Vd}$, $\Np{\Zd}$, and $\Np{\MX}$ for the various fields can also be found in~\eqref{eq:NVZ} and~\eqref{eq:NormCon}, respectively.

The boundary terms in $S_\text{bdy}[\MX,\Vd,\Zd]$ are quite complicated in general, but simplify considerably when evaluated asymptotically at large $r$.  They imply that these fields must be quantized with Neumann boundary conditions for the purposes of computing the generating function of correlation functions.

We will, however, be interested in computing an off-shell Wilsonian object, the Wilsonian influence functional introduced in~\cite{Ghosh:2020lel}, which is the boundary effective action for the hydrodynamic modes. This influence functional is parameterized in terms of the boundary values of the fields $\{\MX_\eta,\Vd,\Zd\}$ (and not their sources), which we denote  as $\{\POp_\eta, \QOp, \EOp\}$.

The Wilsonian influence functional is obtained from the generating function of correlators by a Legendre transform. The latter operation, however, simply removes the boundary term that imposes the Neumann boundary condition. As a consequence, we compute the Wilsonian influence functional directly using the Dirichlet boundary conditions for these fields. These nuances are explained in~\cite{He:2021jna,He:2022deg}, where the reader can find further detail. 

\paragraph{Interaction vertices:} Having parameterized the perturbations of the EMCS system in terms of decoupled fields corresponding to the hydrodynamic modes, we wish to examine the interaction between them. As one might expect, the overall form of the interactions is quite involved as it gets contributions from the Einstein-Maxwell terms in the action~\eqref{eq:SEMax}. One will find terms of the form $\Vd^a\,\Zd^{3-a}$, with $a\in\{0,\ldots,3\}$, as well as parity-preserving terms $\MX_\eta\,\MX_\eta\, \Vd$ and $\MX_\eta\,\MX_\eta\,\Zd$. We will not be interested in these, since they are insensitive to the anomaly. 

Our primary interest lies in the interactions arising from the Chern-Simons term in the bulk. At linear order in $\kcs$ this gives the anomalous non-Gaussian contribution to hydrodynamics. Structurally, this term couples the vector modes of either chirality (or parity), and leads to cubic couplings of the form $\MX_{\eta_1}\,\MX_{\eta_2}\,\MX_{\eta_3}$,  $\MX_{\eta_1} \,\MX_{\eta_2}\, \mathcal{S}$, and $\MX_{\eta_1}\,\mathcal{S}\, \mathcal{S}$, where $\mathcal{S} \in \{\Vd,\Zd\}$. These interactions capture the non-trivial influence functions associated with the anomaly, we focus on them exclusively. 

Plugging in the gauge field parameterization for the perturbations~\eqref{eq:Amaxhydro} into the Chern-Simons term, one finds after some algebra the desired interaction vertices. We organize these by the mixing between the vector and scalar perturbations. The actions are written as integrals over the radial coordinate in the bulk, but in Fourier domain along the boundary in the expressions below. We again emphasize that we have isolated the purely hydrodynamic part of the vertex. In particular, there is an additional contribution involving the transverse photon degrees of freedom $\MY$, which we do not include in our discussion (it is, however, straightforward to obtain given the data in~\cref{sec:vectors}).

In writing the vertices we will employ a set of chiral form factors
$\VVV^{\eta_1\eta_2\eta_3}(\sk_1,\sk_2,\sk_3)$, $\VVSa^{\eta_1\eta_2}(\sk_1,\sk_2,\sk_3)$, and $\VSSa^{\eta}(\sk_1,\sk_2,\sk_3) $, respectively. These are functions of spatial momenta and the chirality label, as indicated, and are defined in~\eqref{eq:VVVffactor},~\eqref{eq:VVSffactor}, and~\eqref{eq:VSSffactor}, respectively. They originate from the fact that the Chern-Simons term induces a cubic pairing involving the  $\epsilon$-symbol in the spatial directions ($\mathbb{R}^3 \subset \mathbb{R} ^{3,1}$ along the boundary).

\begin{itemize}[wide,left=0pt]
\item The simplest vertex corresponds to the self- interaction of the vectorial shear modes. We find a non-trivial vertex for all possible permutations of the chiral polarization index. The vertex, however, can be succinctly presented as
\begin{equation}\label{eq:CSvvv}
\begin{split}
S_{_\text{CS}}^\text{vvv} 
&=  
	\int\, dr\, \int_{\sk_1,\sk_2,\sk_3} \,  \mathbb{N}^{\eta_1\eta_2\eta_3}  
 	 \, \frac{\MX_{\eta_1}(\sk_1)}{r^2} \, 
	\frac{\MX_{\eta_2}(\sk_2)}{r^2}  \,\dv{r}(\frac{\MX_{\eta_3}(\sk_3)}{r^2})  \,,\\
\mathbb{N}^{\eta_1\eta_2\eta_3} &=
 	\frac{2\,i\,\kcs\,\RQ^6}{3\,\mu^3} \, 
 	\bqt_1^2\,\bqt_2^2\,\bqt_3^2\,\omega_2\;
 	\VVV^{\eta_1\eta_2\eta_3}(\sk_1,\sk_2,\sk_3)
\end{split}
\end{equation}	

\item The coupling between two vector modes and a scalar mode has three different structures involving different combinations of the scalar excitations. We find it convenient to write this in terms of the field $\MV$ which is a functional linear combination of the fields dual to the energy transport and charge diffusion, cf.,~\eqref{eq:MZVdiagonal}. It takes the form
\begin{equation}\label{eq:CSvvs}
\begin{split}
S_{_\text{CS}}^\text{vvs} 
&=  \frac{2\,\kcs\,\RQ^4}{3\,\mu^2} \, 
	\int\, dr\, \int_{\sk_1,\sk_2,\sk_3}  
	\left( A^{(1)}_{\eta_1\eta_2}\,\VVSa^{\eta_1\eta_2}(\sk_3,\sk_1,\sk_2) 
	- A^{(2)}_{\eta_1\eta_2}\,\VVSa^{\eta_1\eta_2}(\sk_1,\sk_2,\sk_3) \right) \\
A^{(1)}_{\eta_1\eta_2}
&=
	\frac{\bqt_1^2\,\bqt_2^2}{r^3}\,\MX_{\eta_1}(\sk_1)\bigg(\MX_{\eta_2}(\sk_2)  \left[\frac{1}{r} \,
	\dv{r}(rf\,\dv{r}) - \frac{2i\,\omega_3}{r^2} \dv{r} +\frac{i\omega_3}{r^3}\right] \MV(\sk_3)\\
&\hspace{2.2cm}	
	+  \Dz_+ \MV(\sk_3) \dv{r}(\frac{\MX_{\eta_2}(\sk_2)}{r^2}) -i\omega_2 \dv{\MV(\sk_3)}{r} \frac{\MX_{\eta_2}(k_2)}{r^2} \bigg) + (\sk_2 \leftrightarrow \sk_3)  \,,\\
A^{(2)}_{\eta_1\eta_2}
&= \frac{2 \,\bqt_2^2\,\bqt_3^2}{r^3}\,  \MX_{\eta_2}(\sk_3) 
	\left[ \Dz_+ \MV(k_1) \dv{r}(\frac{\MX_{\eta_1}(k_2)}{r^2}) -i\omega_2 \dv{\MV(k_1)}{r} \frac{\MX_{\eta_1}(k_2)}{r^2} \right].
\end{split}
\end{equation}
\item The coupling of a vector mode to two scalar modes is again quite simple when expressed in terms of $\MV$. One finds 
\begin{equation}\label{eq:CSvss}
\begin{split}
S_{_\text{CS}}^\text{vss}
&=
	\frac{8i\,\kcs\,\RQ^2}{3\,\mu} \,	\int \, dr \int_{\sk_1,\sk_2,\sk_3}  
	\omega_1\,\bqt_3^2 \, 
	   \frac{\MV(\sk_1)}{r^2}\,\dv{\MV(\sk_2)}{r}\, \frac{\MX_{\eta}(\sk_3)}{r^2} \VSSa^{\eta}(\sk_1,\sk_2,\sk_3)\,.
\end{split}
\end{equation}
\end{itemize}
This completes the data we need from the bulk for evaluating the effective action for the hydrodynamic modes. 

\section{The anomalous effective action}\label{sec:anomalySeff}

Varying the bulk action~\eqref{eq:XVZbulk} for the linearized fluctuations $\{\MX_\eta,\Vd,\Zd\}$ we obtain the linearized equations for the modes. Of interest is the solution to these equations in the domain of outer communication of the background geometry~\eqref{eq:RNAdS}, i.e., in the region $r\in (r_+,\infty)$. The strategy, as explained in~\cite{Jana:2020vyx}, is to solve for the ingoing boundary-bulk propagator $\Gin{}(r,\sk)$ for the fields in question. This propagator satisfies regularity at the future horizon, and has a non-normalizable fall-off at the boundary. We conveniently choose the coefficient of the non-normalizable mode to be unity. The ingoing Green's functions are themselves obtained in a gradient expansion in powers of $\bwt$ and $\bqt$. The expressions for the scalar sector fields $\Vd$ and $\Zd$ have, as noted, already been obtained in~\cite{He:2022deg}, which we directly import for our analysis. The propagator for $\MX_\eta$ is derived in~\cref{sec:Xsols}. For all the fields we have explicit expressions for the ingoing Green's function to quartic order in the $\bwt$, $\bqt$ gradient expansion.  

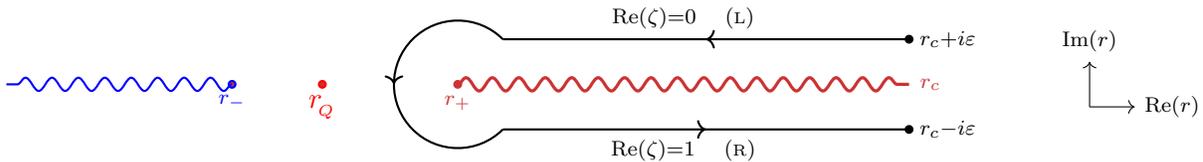
\begin{figure}[h!]
\begin{center}
\begin{tikzpicture}[scale=0.6]
\draw[thick,color=rust,fill=rust] (-5,0) circle (0.45ex);
\draw[thick,color=black,fill=black] (5,1) circle (0.45ex);
\draw[thick,color=black,fill=black] (5,-1) circle (0.45ex);
\draw[very thick,snake it, color=rust] (-5,0) node [below] {$\scriptstyle{r_+}$} -- (5,0) node [right] {$\scriptstyle{r_c}$};
\draw[thick,color=black, ->-] (5,1)  node [right] {$\scriptstyle{r_c+i\varepsilon}$} -- (0,1) node [above] {$\scriptstyle{\Re(\ctor) =0}\quad (\skL) $} -- (-4,1);
\draw[thick,color=black,->-] (-4,-1) -- (0,-1) node [below] {$\scriptstyle{\Re(\ctor) =1}\quad (\skR) $} -- (5,-1) node [right] {$\scriptstyle{r_c-i\varepsilon}$};
\draw[thick,color=black,->-] (-4,1) arc (45:315:1.414);
\draw[thin, color=black,  ->] (9,-0.5) -- (9,0.5) node [above] {$\scriptstyle{\Im(r)}$};
\draw[thin, color=black,  ->] (9,-0.5) -- (10,-0.5) node [right] {$\scriptstyle{\Re(r)}$};  
\draw[thick,color=blue,fill=rust] (-10,0) circle (0.45ex);
\draw[thick,color=red,fill=rust] (-8,0) circle (0.45ex) node[below] {$\RQ$};
\draw[thick,snake it, color=blue] (-10,0) node [below] {$\scriptstyle{r_-}$} -- (-15,0) ;
\end{tikzpicture}
\caption{ The complex $r$ plane with the locations of the two regulated boundaries (with cut-off $r_c$), the outer and inner horizons at $r_\pm$ and the radius $\RQ$ defined in~\eqref{eq:hLkrqC} marked. The grSK contour is a codimension-1 surface in this plane (drawn at fixed $v$). As indicated the direction of the contour is counter-clockwise. It encircles the branch point at the outer horizon with the cut running out to the boundary. The cut emanating from the inner horizon is in particular oriented away, depicted (as are those from the other roots of $f(r)$, which we have not displayed).}
\label{fig:mockt}
\end{center}
\end{figure}

The knowledge of the ingoing boundary-bulk Green's functions suffices to compute any Schwinger-Keldysh correlation function. To do so, we first extend the solution to the complex radial domain and map the boundary Schwinger-Keldysh contour, to a contour in this complexified radial plane~\cite{Glorioso:2018mmw}. Following,~\cite{Jana:2020vyx} we find it convenient to introduce $\ctor(r)$ using
\begin{equation}\label{eq:mockT}
\dv{\ctor}{r} = \frac{2}{i\,\beta}\, \frac{1}{r^2f}\,.
\end{equation}	
This coordinate is normalized such that $\lim_{r\to \infty+i\epsilon}\, \Re(\ctor) =0$ and $\lim_{r\to \infty-i\epsilon}\,\Re(\ctor ) =1$. These two boundaries correspond to the L and R segments of the boundary Schwinger-Keldysh contour, respectively, as depicted in~\cref{fig:mockt}. For the most part we will not need an explicit functional form for $\ctor$, but it is easy to obtain. We give an expression in~\cref{sec:mockT} with the correct orientation of the cuts for $\ctor(r)$. 

Expressing the ingoing Green's function as $\Gin{}(\ctor,\sk)$, the outgoing Green's function can be obtained using the time-reversal symmetry of the background geometry. For time-reversal invariant dynamical systems. Likewise,~\cref{eq:XVZbulk}, one has $\Gout{}(\ctor,\sk) = e^{-\beta\omega\,\ctor}\,\Gin{}(\ctor,\bar{\sk})$. We recall that $\bar{\sk}^\mu = (-\omega,\vb{k})$ is the frequency reversed four-momentum. 
Consequently, the solution of the linear wave equation for a field $\mathsf{M} \in \{\MX_\eta, \Zd, \Vd\}$ on the complexified spacetime, referred to as the grSK geometry, is then 
\begin{equation}\label{eq:phSKsol}
\mathsf{M}(\ctor, \sk)  = -\Gin{\mathsf{M}}(\ctor,\sk)\, \JF + 
\Gout{\mathsf{M}}(\ctor,\sk)\, e^{\beta\omega} \, \JP\,.
\end{equation}	
Here $\JF$ and $\JP$ are the retarded and advanced combinations of the Schwinger-Keldysh L and R sources, dressed with statistical factors ($\nB$) designed to simplify the KMS relations
\begin{equation}\label{eq:JFP}
\JF = -(1+ \nB)\, J_\skR  + \nB\, J_\skL \,, \qquad \JP = -\nB \, (J_\skR - J_\skL)\,, 
\qquad \nB(\omega) \equiv \frac{1}{e^{\beta\omega}-1}\,.
\end{equation}	
Furthermore, the information of the ingoing boundary-bulk propagator suffices to also fix the bulk-bulk propagator~\cite{Loganayagam:2022zmq}. Thus, once one has derived $\Gin{}(r,\sk)$, one can compute Witten diagrams on the grSK geometry, which would give us the generating function of the hydrodynamic correlators. 

However, we would like to obtain the effective action for the hydrodynamic modes. This is achieved by Legendre transforming the generating function. A-priori the fields $\{\MX_\eta,\Vd, \Zd\}$  are quantized with Neumann boundary conditions. This is a consequence of these fields being related to the physical metric and gauge field perturbation through derivative redefinitions, cf.,~\cref{sec:vectors}. Since the physical metric and gauge field perturbations are quantized with Dirichlet boundary conditions, it follows that the redefined variables $\{\MX_\eta,\Vd, \Zd\}$, which have diagonal kinetic terms, obey Neumann boundary conditions. This has a salubrious effect. As explained in~\cite{Ghosh:2020lel}, the Legendre transform from the generating function to the Wilsonian influence function is tantamount to quantizing the fields with Dirichlet boundary conditions.  Armed with this information, it is straightforward to evaluate the on-shell action of the bulk theory parameterized  the hydrodynamic fields, which is the physical hydrodynamic effective action we are after. 

The computation of the Wilsonian influence functionals from the grSK geometry in the context of hydrodynamics has been explained in~\cite{Ghosh:2020lel,He:2022jnc,He:2021jna,He:2022deg}. A general discussion including computation of non-Gaussian terms can be found in~\cite{Loganayagam:2022zmq}. In general, all one needs to do is to adopt the standard Witten diagrammatics to the grSK contour, and compute the vertices in the effective action to the desired order. 

When the dust settles the hydrodynamic effective action we are after takes the form
\begin{equation}\label{eq:SVZall}
\frac{8\pi^2}{N^2} \, S\left[\POp_\eta,\QOp,\EOp\right] 
	= S_{\text{contact}}\left[\QOp, \EOp\right] + S_{_\text{WIF}}^{(2)}\left[\POp_\eta,\QOp,\EOp\right] + + S_{_\text{WIF}}^{(3)}\left[\POp_\eta,\QOp,\EOp\right] + \cdots \,.
\end{equation}
The contact term is the Class L action of a charged fluid (Legendre transformed to be written in terms of the fields). It exists because the background solution has non-trivial free energy density. We will not elaborate on its contribution further here, but refer the reader to~\cite{He:2021jna,He:2022deg} for a detailed analysis of this term. We describe the general features of the Gaussian part, and the anomaly constrained piece of the cubic part in~\cref{sec:s2wif} and~\cref{sec:s3wif}, respectively.

\subsection{The Gaussian effective action}
\label{sec:s2wif}

\begin{figure}[t]
\subfloat{
\begin{minipage}[h!]{0.45\textwidth}
\begin{tikzpicture}
  \node (img)  {\includegraphics[scale=0.75]{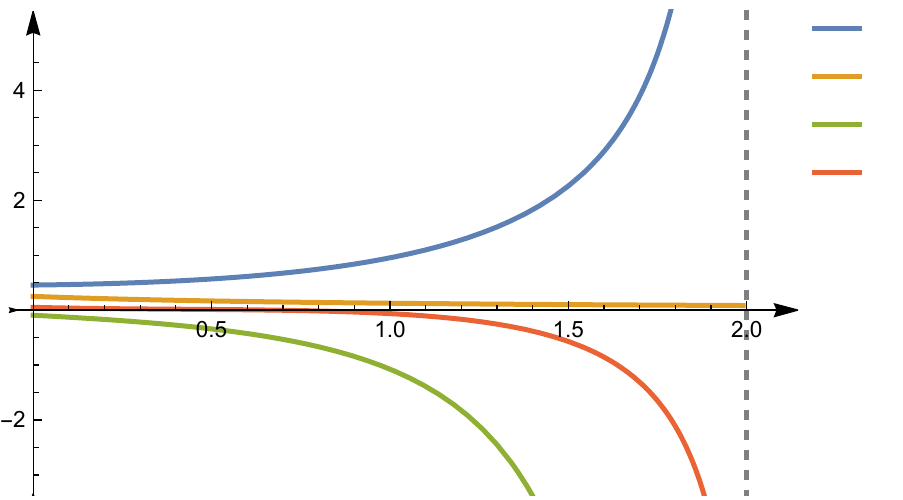}};
  \node[right=of img, node distance=0cm,font=\small,yshift=-0.5cm,xshift=-2cm] {$Q^2$};
  \node[right=of img, node distance=0cm, anchor=center,font=\scriptsize,yshift=1.8cm,xshift=-1cm] {$\mathfrak{h}^{\scriptstyle{s,\eta}}_{3,0}$};
  \node[right=of img, node distance=0cm, anchor=center,font=\scriptsize,yshift=1.35cm,xshift=-1cm] {$\mathfrak{h}^{\scriptstyle{s,\eta}}_{1,2}$};
  \node[right=of img, node distance=0cm, anchor=center,font=\scriptsize,yshift=0.95cm,xshift=-1cm] {$\mathfrak{h}^{\scriptstyle{s,\eta}}_{4,0}$};
  \node[right=of img, node distance=0cm, anchor=center,font=\scriptsize,yshift=0.55cm,xshift=-1cm] {$\mathfrak{h}^{\scriptstyle{s,\eta}}_{2,2}$};
 \end{tikzpicture}
 \end{minipage}}
 \subfloat{
\begin{minipage}[h!]{0.45\textwidth}
\hspace{0.7cm}
\begin{tikzpicture}
  \node (img)  {\includegraphics[scale=0.35]{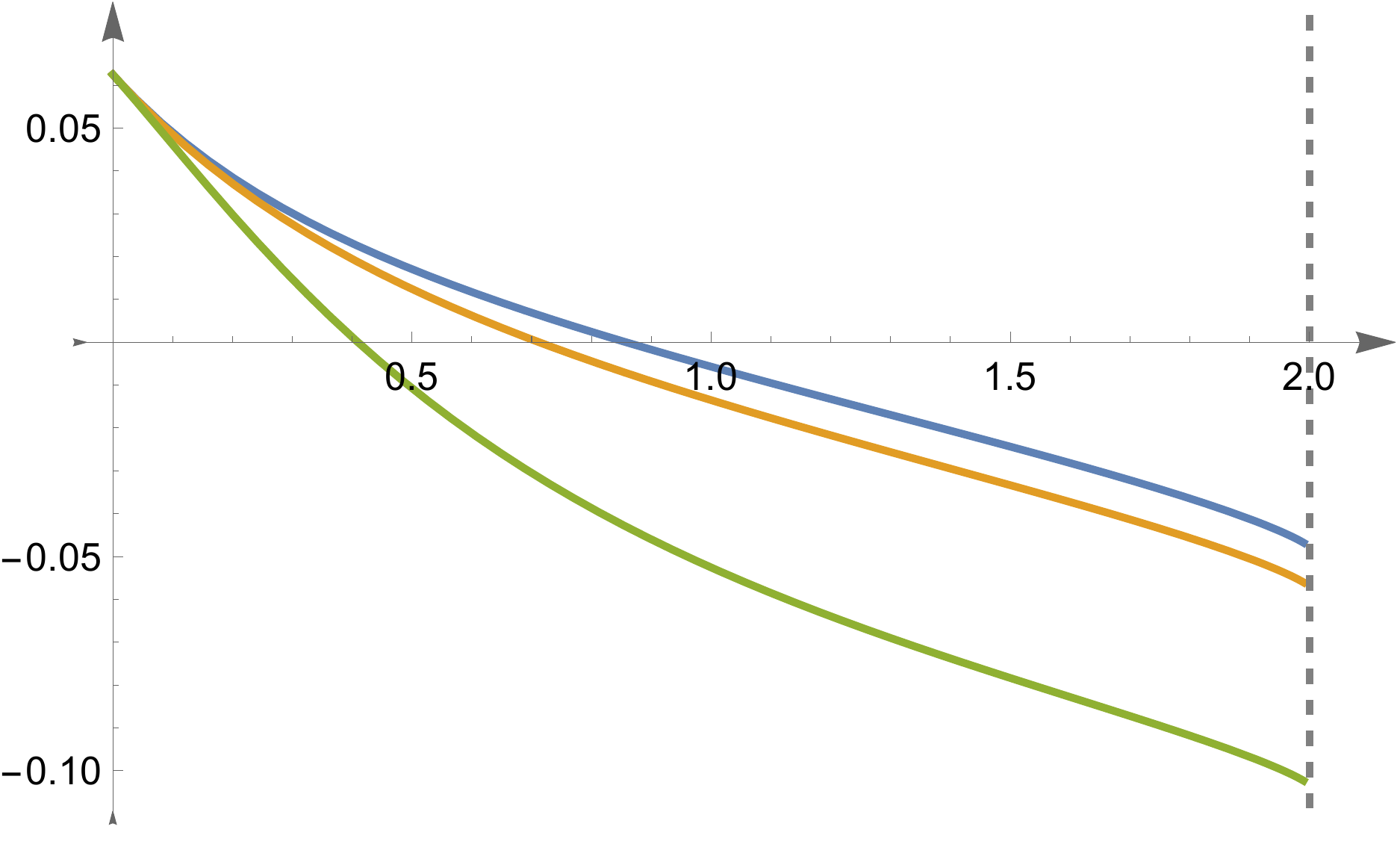}};
  \node[below=of img, node distance=0cm,font=\small,yshift=3.8cm,xshift=3.6cm] {$Q^2$};
  \node[left=of img, node distance=0cm, anchor=center,font=\small,yshift=2.3cm,xshift=1.7cm] {$\mathfrak{h}^{\scriptstyle{s,\eta}}_{0,4}$};
  \node[below=of img, node distance=0cm,font=\scriptsize,yshift=3cm,xshift=3.6cm] {$\kcs=0$};
  \node[below=of img, node distance=0cm,font=\scriptsize,yshift=2.6cm,xshift=3.8cm] {$\kcs=\frac{1}{2\sqrt{6}}$};
  \node[below=of img, node distance=0cm,font=\scriptsize,yshift=1.8cm,xshift=3.6cm] {$\kcs=\frac{1}{2}$};
 \end{tikzpicture}
 \end{minipage}}
 \caption{The coefficients at cubic and quartic order for the shear dispersion function $\Kin{\MX,\eta}(\bwt,\bqt)$. Of the coefficients $\mathfrak{h}_{m,n}^{s,\eta}$ only $\mathfrak{h}_{0,4}^{s,\eta}$ depends on the Chern-Simons coefficient (and therefore on the chirality label $\eta$).}
 \label{fig:KxCoeffs}
 \end{figure}
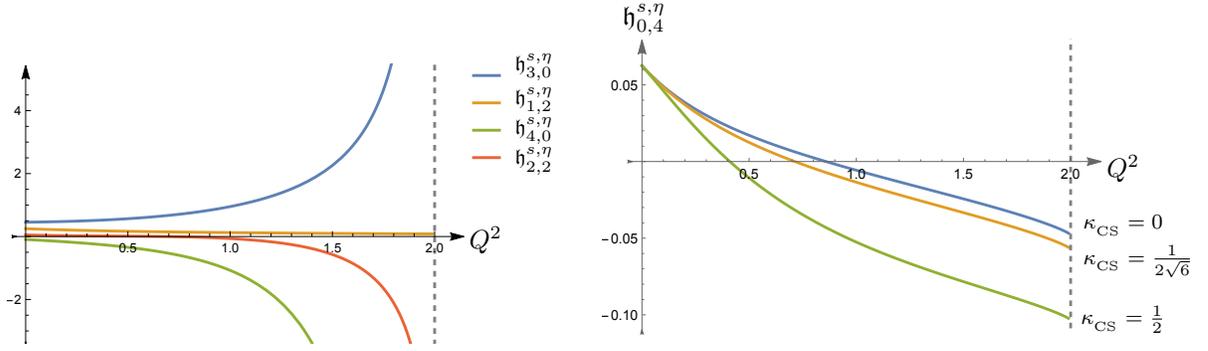

\begin{figure}[ht]
\subfloat{
\begin{minipage}[h!]{0.5\textwidth}
\hspace{3cm}
\begin{tikzpicture}
  \node (img)  {\includegraphics[scale=1]{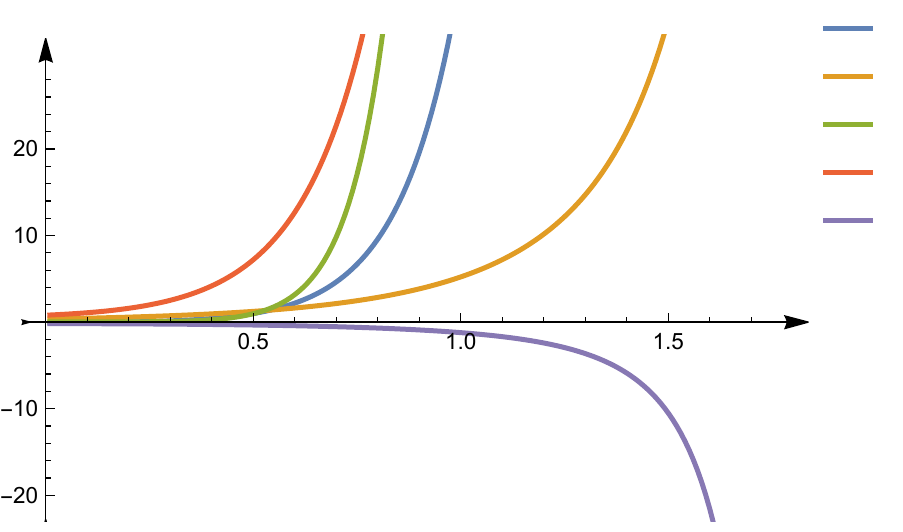}};
  \node[right=of img, node distance=0cm,font=\small,yshift=-0.5cm,xshift=-2cm] {$Q^2$};
  \node[right=of img, node distance=0cm, anchor=center,font=\scriptsize,yshift=2.4cm,xshift=-1cm] {$\mathfrak{h}^{\scriptstyle{c}}_{3,0}$};
  \node[right=of img, node distance=0cm, anchor=center,font=\scriptsize,yshift=1.9cm,xshift=-1cm] {$\mathfrak{h}^{\scriptstyle{c}}_{1,2}$};
  \node[right=of img, node distance=0cm, anchor=center,font=\scriptsize,yshift=1.4cm,xshift=-1cm] {$\mathfrak{h}^{\scriptstyle{c}}_{4,0}$};
  \node[right=of img, node distance=0cm, anchor=center,font=\scriptsize,yshift=0.9cm,xshift=-1cm] {$\mathfrak{h}^{\scriptstyle{c}}_{2,2}$};
  \node[right=of img, node distance=0cm, anchor=center,font=\scriptsize,yshift=0.4cm,xshift=-1cm] {$\mathfrak{h}^{\scriptstyle{c}}_{0,4}$};
 \end{tikzpicture}
 \end{minipage}}
 \caption{The coefficients at cubic and quartic order for the shear dispersion function $K_c(\bwt,\bqt)$. These coefficients diverge quite rapidly as $Q^2 \to 2$, the extremal value. }
 \label{fig:KcCoeffs}
 \end{figure}
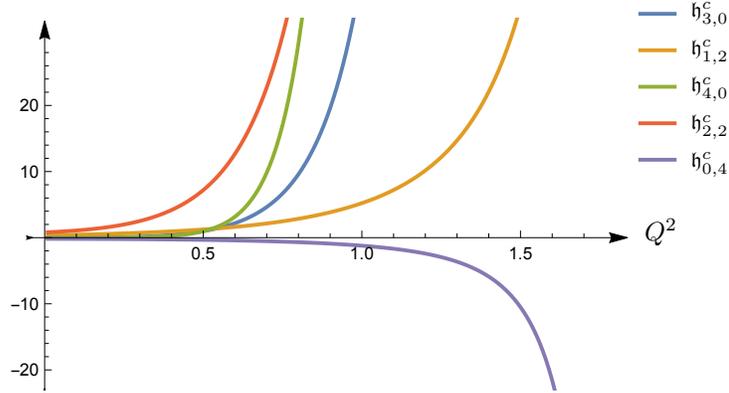

The dynamical Wilsonian influence functional characterizes the physical fluctuations of the charged plasma. As indicated, it  has Gaussian and non-Gaussian contributions, with the Gaussian part nicely decoupled into the chiral shear, charge diffusion, and sound modes as noted in~\eqref{eq:SanomA}. In the advanced retarded basis the Gaussian part of the action is simply
\begin{equation}\label{eq:WIF2}
\begin{split}
S_{_\text{WIF}}^{(2)} 
&=  - \int_\sk\,  \frac{k^2}{\nB(-\omega)} \bigg[ 
		\POp^\eta_{_\text{P}}(-\sk)\, \Kin{\text{shear},\eta}(\sk)\, \POp^\eta_{_\text{F}}(\sk) 
		+ \QOp_{_\text{P}}(-\sk)\, \Kin{\text{diff}}(k)\, \QOp_{_\text{F}}(\sk)
		 \\
&  \hspace{3.0cm} 
		+\; \EOp_{_\text{P}}(-\sk)\, \Kin{\text{sound}}(\sk)\, \EOp_{_\text{F}}(\sk) \bigg]  
\end{split}
\end{equation}	
This is just a rewriting of~\eqref{eq:SanomA} in the more convenient advanced/retarded basis~\cref{eq:JFP}. Analytic expressions for the dispersion relations, which are the inverse hydrodynamic Green's functions, can be  obtained in a gradient expansion in frequencies and momenta. They take the form 
\begin{equation}\label{eq:DispKer}
\begin{split}
\Kin{\text{shear},\eta}(\omega,\bk)
&= 
	\Np{\MX,\eta}(\BQTCS)\,\Kin{\MX,\eta}(\bwt,\bqt)  \\
\Kin{\text{diff}}(\omega,\bk)
&= 
	\frac{\Np{\Vd}(\BQT)\, r_+^4}{(r_+^2-\RQ^2)^2} \, K_c(\bwt,\bqt) \\ 
\Kin{\text{sound}}(\omega,\bk)
&=
	   \frac{\Np{\Zd}(\BQT)}{72\, (1+Q^2)\,r_+^2}\, K_s(\bwt,\bqt)  \,, 
\end{split}
\end{equation}
The dispersion functions for the chiral shear waves and charge diffusion, are diffusive, and have a gradient expansion of the form
\begin{equation}\label{eq:hdispcfs}
\begin{split}
\Kin{\MX,\eta}(\bwt,\bqt) 
&= 
	-i\,\bwt + \frac{1}{4}\, \left(1-\frac{2\,\RQ^2}{3\,r_+^2}\right))\bqt^2 - \Dfn{3}{2,0}(r_+)\, \bwt^2  
	+ \sum_{\substack{m,n=1 \\ n+m\geq 3}}^\infty\, (-i)^m\, \mathfrak{h}^{\scriptstyle{s,\eta}}_{m,n}\, \bwt^m\, \bqt^n   \\ 
K_c(\bwt,\bqt)
&=
	-i\,\bwt + \frac{1}{2}\, \left(1-\frac{\RQ^2}{3\,r_+^2}\right)\bqt^2
	 - \Dfn{\MY}{2,0}(r_+)\, \bwt^2  +\sum_{\substack{n,m=1 \\ n+m\geq 3}}^\infty\,  (-i)^m\,\mathfrak{h}^{\scriptstyle{c}}_{m,n}\, \bwt^m\, \bqt^n 
\end{split}
\end{equation}
We have written out the explicit coefficients as functions of $r_+$ and $Q$ (which, we recall, stand as proxy for the temperature and chemical potential, cf.,~\eqref{eq:TRN}) in terms where the expressions are not too complicated. For instance, the leading coefficients $\Dfn{3}{2,0}(r_+)$ and $\Dfn{\MY}{2,0}(r_+)$ are, respectively, given by
\begin{equation}\label{eq:Dfnsanalytic}
\begin{split}
\Dfn{3}{2,0}(r_+) 
&= 
	\frac{1}{2} + \frac{1+Q^2}{2\,\sqrt{1+4\,Q^2}}\, 
	\log\left(\frac{3-\sqrt{1+4\,Q^2}}{3+\sqrt{1+4\,Q^2}}\right), \\
\Dfn{\MY}{2,0}(r_+) 
&= 
	\frac{1}{4\,h(r_+)^2(1+4\,Q^2)}\left[-6\,\frac{\RQ^2}{r_+^2}h(r_+)+(1+4\,Q^2)\log\left(2-Q^2\right)\right. \\ 
	&\quad 
	\left.
	+\left(h(r_+)-\frac{\RQ^2}{r_+^2}\left(1-10\, \frac{\RQ^2}{r_+^2}\right)\right)\frac{(1+Q^2)}{\sqrt{1+4\,Q^2}}\log\left(\frac{3-\sqrt{1+4\,Q^2}}{3+\sqrt{1+4\,Q^2}}\right)\right].	
\end{split}
\end{equation}
However, at higher orders the coefficients have involved analytic expressions owing to the charge dependence. We have therefore opted to illustrate the behaviour graphically in~\cref{fig:KxCoeffs,fig:KcCoeffs}.  For completeness, we include  certain analytic expressions in terms of the functions appearing in the bulk solution in~\cref{sec:dispersionfns}.

On the other hand,  $K_s(\bwt,\bqt)$ has a propagating part damped by shear attenuation, as appropriate for sound dispersion. It is in fact quite simple to quartic order (expressions to higher orders can be found in~\cite{He:2022deg})
\begin{equation}
K_s(\bwt,\bqt) = -\bwt^2 + \frac{\bqt^2}{3} + \frac{1}{3}\, \frac{\bqt^2}{1+Q^2}\left(-i\,\bwt + \Dfn{2,0}{3}(r_+) \,\bwt^2 + \frac{\bqt^2}{6}  + \frac{\RQ^2}{2\,r_+^2}\, \left(\bwt^2 - \frac{\bqt^2}{3}\right)\right)  .
\end{equation}	

The vanishing locus of the inverse Green's functions $\Kin{}$ above, gives us the physical dispersion relations for the low-lying hydrodynamic modes. Solving for $\bwt(\bqt)$ one finds
\begin{itemize}[wide,left=0pt]
\item Chiral momentum diffusion modes in the vector sector, with dispersion
\begin{equation}\label{eq:Mdiff}
\bwt_{_{\text{shear},\eta}} = -\,\frac{i\, \bqt^2}{4\,(1+Q^2)} +\eta\, \kcs \sqrt{\frac{3}{2}}\, \frac{Q^3}{2 (1+Q^2)} \bqt^3 + \order{\bqt^4} \,.
\end{equation}	
\item A charge diffusion mode insensitive to parity breaking with dispersion
\begin{equation}\label{eq:Qdiff}
\bwt_{_\text{diff }} = - i\frac{2+Q^2}{4\,(1+Q^2)}\, \bqt^2 + \order{\bqt^4} \,.
\end{equation}	
\item A pair of sound modes also oblivious to parity breaking with dispersion as computed in
\begin{equation}\label{eq:soundw}
\bwt_{_\text{sound}} 
= 
	\pm \frac{\bqt}{\sqrt{3}}-\frac{i\, \bqt^2}{6\,(1+Q^2)} \pm \frac{1}{6\sqrt{3}\, (1+Q^2)^2} \left[\frac{1}{4}+ \frac{1}{2}\, Q^2 +  (1+Q^2)\, \Delta_{2,0}(r_+)\right] \bqt^3 + \order{\bqt^4}\,.
\end{equation}
\end{itemize}

The parity even dispersions are just special cases of the results derived in~\cite{He:2022deg}. The parity odd result for the chiral shear waves is consistent with previous derivations in the literature. This result was first obtained from an analogous linearized gravity computation in ~\cite{Sahoo:2009yq}. We have checked that the results we quoted are consistent with the 
solution of the conservation equations for the second order stress tensor and charge current obtained using the non-linear fluid/gravity solution in~\cite{Banerjee:2008th,Erdmenger:2008rm}. 

These investigations found three first order transport coefficients: (a) shear viscosity, which captures the shear damping rate, (b) charge conductivity, which captures the charge diffusion rate, and (c) a parity odd anomaly constrained coefficient in the charge current. The latter two are consistent with the $\bqt^2$ term in the corresponding dispersion relation, while the parity-odd coefficient does not contribute to the linearized equations. 

At second order, there are 9 transport coefficients in the stress tensor (two identically vanish), and 5 coefficients in the charge current. Of these only two, one each in the stress tensor and charge current, are linear in the anomaly coefficient. The term in the charge current drops out of the linearized fluctuations, so the only anomaly constrained coefficient that can be deduced from the dispersion relations is the second order coefficient in the stress tensor. The tensor structure involved is the symmetrized Weyl covariant derivative of the vorticity vector $\ell^\mu = \epsilon^{\nu\lambda \sigma \mu} \, u_\nu \, \nabla_\lambda u_\sigma$. The transport coefficient, which we can deduce from~\eqref{eq:Mdiff} turns out to be $- \kcs\,\frac{\sqrt{6}\, Q^3}{1+Q^2}\, r_+^2 $, and agrees with that obtained in these earlier works.\footnote{
 The papers~\cite{Banerjee:2008th} and~\cite{Erdmenger:2008rm} use differing normalization for the EMCS dynamics. Moreover, they both employ conventions different from what we chose. Nevertheless, upon suitably translating through the convention choices, we find that we reproduce the transport data quoted in these works. The parity odd coefficient that is captured by the chiral shear dispersion is denoted $\mathcal{N}_{7}$ in ~\cite{Banerjee:2008th} and $\tilde{\lambda}_2$ in~\cite{Erdmenger:2008rm}.  }

\subsection{Anomalous non-Gaussian effective action} 
\label{sec:s3wif}

The non-Gaussian contributions to the effective action at cubic order have both parity-even and parity-odd components. The former are terms that exist for any charged plasma, while the latter are anomaly constrained.  As explained earlier, they originate from the bulk Chern-Simons term and are linear in $\kcs$ (which fixes the anomaly coefficient $c_A$). The bulk dynamics predicts cubic couplings between the chiral vector modes, and the charge diffusion and sound modes,~\eqref{eq:CSvvv}-\eqref{eq:CSvss}. The cubic influence functional has couplings fixed by the three-point functions of the fields. 

An advantage of the advanced/retarded basis~\eqref{eq:JFP} is that correlators involving solely advanced or solely retarded fields vanishes identically. In the field theory this is a consequence of the Schwinger-Keldysh and KMS conditions on thermal correlators, both of which are implemented in the grSK construction by the complex radial contour prescription.  Thus, of the various combinations of retarded and advanced fields, the only non-vanishing correlation functions are ones with 2 retarded and one advanced, or 2 advanced and one retarded field, viz., $I_{_\text{FFP}}(\sk_1,\sk_2,\sk_3)$ and $I_{_\text{PPF}}(\sk_1,\sk_2,\sk_3)$. 

From the Chern-Simons couplings, we have cubic interactions between 3 elements of the set $\{\POp^\eta,\QOp,\EOp\}$, with terms constrained to contain at least one $\POp^\eta$. Additionally, we should also account for 
assignment of P/F labels, and for permutations of the fields. Therefore, in the Wilsonian influence functional,  we will have a myriad set of terms.

Consider for simplicity a cubic vertex in the bulk coupling three fields $\mathsf{M}_i$ for $i=1,2,3$,
\begin{equation}\label{eq:vertexM}
\int dr \int_{\sk_1,\sk_2,\sk_3} \, \mathfrak{v}\left(r,\dv{r},\sk_1,\sk_2,\sk_3\right)
	\mathsf{M}_1(r,\sk_1)\,\mathsf{M}_2(r,\sk_2)\,\mathsf{M}_3(r,\sk_3).
\end{equation}	
We  are assuming that  $\mathsf{M}_i \in \{\MX_\eta,\Vd,\Zd\}$. Note also that the vertex in the bulk may involve derivatives acting on the fields $\mathsf{M}_i$, as can be seen from~\eqref{eq:CSvvv}-\eqref{eq:CSvss}. This vertex contributes to the cubic term in the effective action for the dual hydrodynamic modes, denoted $\mathcal{O}_i \in \{\POp^\eta,\QOp,\EOp\}$ in the following manner:
\begin{equation}\label{eq:S3M}
\begin{split}
S_{_\text{WIF}}^{(3)}
&= 
	\int_{\sk_1,\sk_2,\sk_3}\, \bigg[I_{_\text{FFP}}(\sk_1,\sk_2,\sk_3) \, \Kin{\mathcal{O}_1}(\sk_1)\, \Kin{\mathcal{O}_2}(\sk_2)\, \Kin{\mathcal{O}_3}(\bar{\sk}_3) \,  \mathcal{O}_{1,\text{F}}(\sk_1) \, \mathcal{O}_{2,\text{F}}(\sk_2)\, \mathcal{O}_{3,\text{P}}(\sk_3)\\
&\; +
	I_{_\text{PPF}} (\sk_1,\sk_2,\sk_3) \, \Kin{\mathcal{O}_1}(\bar{\sk}_1)\, \Kin{\mathcal{O}_2}(\bar{\sk}_2)\, \Kin{\mathcal{O}_3}(\sk_3)  \, 
	\mathcal{O}_{1,\text{P}}(\sk_1) \, \mathcal{O}_{2,\text{P}}(\sk_2)\, \mathcal{O}_{3,\text{F}}(\sk_3) + \text{perms}\bigg].
\end{split}
\end{equation}
The permutations involve swapping the advanced/retarded fields (P/F), with the advanced fields appearing with the reversed propagator $\Kin{}(\bar{\sk})$. Specifically, given the structure of the bulk interaction terms, we expect therefore the following non-Gaussian terms to be induced by the anomaly
\begin{itemize}[wide,left=0pt]
\item  Cubic vertices involving 3 shear modes arising from~\eqref{eq:CSvvv} with terms of the form
\begin{equation}\label{eq:PPPterms}
	\POp^{\eta_1}_{_\text{F}}(\sk_1) \, \POp^{\eta_2}_{_\text{F}}(\sk_2)\, \POp^{\eta_3}_{_\text{P}}(\sk_3), 
\qquad 
	\POp^{\eta_1}_{_\text{P}}(\sk_1) \, \POp^{\eta_2}_{_\text{P}}(\sk_2)\, \POp^{\eta_3}_{_\text{F}}(\sk_3) \,. 
\end{equation}
\item Letting $\mathcal{S} \in \{\QOp,\EOp\}$, cubic vertices with 2 shear modes from~\eqref{eq:CSvvs} involving
\begin{equation}\label{eq:PPSterms}
\begin{split}
&	 
	\POp^{\eta_1}_{_\text{F}}(\sk_1) \, \POp^{\eta_2}_{_\text{F}}(\sk_2)\, \mathcal{S}_{_\text{P}}(\sk_3),
\quad	
	\POp^{\eta_1}_{_\text{F}}(\sk_1) \, \POp^{\eta_2}_{_\text{P}}(\sk_2)\, \mathcal{S}_{_\text{F}}(\sk_3) ,\\
&
	\POp^{\eta_1}_{_\text{P}}(\sk_1) \, \POp^{\eta_2}_{_\text{P}}(\sk_2)\, \mathcal{S}_{_\text{F}}(\sk_3),
\quad	
		\POp^{\eta_1}_{_\text{P}}(\sk_1) \, \POp^{\eta_2}_{_\text{F}}(\sk_2)\, \mathcal{S}_{_\text{P}}(\sk_3)\,.
\end{split}
\end{equation}
\item Finally, vertices with a single shear mode originating from~\eqref{eq:CSvss}   
\begin{equation}\label{eq:PSSterms}
\begin{split}
&
	\POp^{\eta_1}_{_\text{F}}(\sk_1) \, \mathcal{S}_{_\text{F}}(\sk_2)\, \mathcal{S}_{_\text{P}}(\sk_3) ,
\quad	
	\POp^{\eta_1}_{_\text{P}}(\sk_1) \, \mathcal{S}_{_\text{F}}(\sk_2)\, \mathcal{S}_{_\text{F}}(\sk_3) ,\\
&
	\POp^{\eta_1}_{_\text{P}}(\sk_1) \, \mathcal{S}_{_\text{P}}(\sk_2)\, \mathcal{S}_{_\text{F}}(\sk_3) ,
\quad	
	\POp^{\eta_1}_{_\text{F}}(\sk_1) \, \mathcal{S}_{_\text{P}}(\sk_2)\, \mathcal{S}_{_\text{P}}(\sk_3) \,. 
\end{split}		
\end{equation}
\end{itemize}

A-priori this seems to be a lot of vertices. One expects a more concise presentation of this data when re-expressed in terms of the conserved current correlators. However, parameterizing the current in terms of the hydrodynamic modes leads to this multitude of vertices. For the present we will sketch how to compute such vertices in the effective field theory, and postpone the discussion of efficiently organizing the result to~\cref{sec:discuss}. 

The vertices in the Wilsonian influence functional~\eqref{eq:S3M}, depend on the 3-point function of the corresponding operators,  $I_{_\text{FFP}}$, $I_{_\text{PPF}}$, etc. These 3-point functions can be computed using bulk Witten diagrams, which are evaluated on the contour depicted in~\cref{fig:mockt}. Given a bulk vertex of the form~\eqref{eq:vertexM}, the contact Witten diagram is computed by convolving the bulk-boundary propagators with the specified vertex function. The contour integral prescription for contact diagrams amounts to picking up the discontinuity across the horizon branch cut, and thus leads to an integral over the region outside the horizon, cf.,~\cite{Jana:2020vyx,Loganayagam:2022zmq}. For example, the result for the FFP and PPF correlators can be expressed (modulo a caveat discussed below) as follows:
\begin{equation}\label{eq:IFFPintegrals}
\begin{split}
I_{_\text{FFP}} 
&= 
	(1-e^{\beta\omega_3})\, \int_{r_+}^\infty\, dr\, \mathfrak{v}\left(r,\dv{r},\sk_1,\sk_2,\sk_3\right)\Gin{\mathsf{M}_1}(\ctor,\sk_1)\, \Gin{\mathsf{M}_2}(\ctor,\sk_2) \, e^{-\beta\omega_3\,\ctor}\, \Gin{\mathsf{M}_3}(\ctor,\bar{\sk}_3) \,,\\
I_{_\text{PPF}} 
&= 
	(1-e^{-\beta\omega_3})\, \int_{r_+}^\infty\, dr\, \mathfrak{v}\left(r,\dv{r},\sk_1,\sk_2,\sk_3\right)  e^{-\beta(\omega_1+\omega_2)\,\ctor}
	\Gin{\mathsf{M}_1}(\ctor,\bar{\sk}_1)\, \Gin{\mathsf{M}_2}(\ctor,\bar{\sk}_2) \,\, \Gin{\mathsf{M}_3}(\ctor,\sk_3) \,.
\end{split}
\end{equation}	
In writing the second line we have employed  energy conservation of the interaction vertex  to rewrite $\omega_1+\omega_2 = -\omega_3$ (in the prefactor). We caution the reader that our compact notation for the vertex function $\mathfrak{v}$ hides the fact that some of these boundary-bulk propagators, and the radial Boltzmann weights $e^{-\beta\omega\ctor}$, might be acted upon by derivative operators. 

With this understanding we can compute any of the 3-point functions which originate from the bulk Chern-Simons term. Note that  boundary-bulk Green's functions for the fields have the following divergent asymptotic behaviour:
\begin{equation}\label{eq:GinasymXVZ}
\begin{split}
\Gin{\MX,\eta}(r,\sk) 
&\to 
	 \frac{1}{2}\, \Kin{\MX,\eta}(\sk)\,r^2 + \order{r^0}\,, \\
\Gin{\Vd}(r,\sk) 
&\to 
	 \frac{r_+^4\, K_c(\sk)}{(r_+^2-\RQ^2)^2} \,  \log r+\order{r^0}\,, \\
\Gin{\Zd}(r,\sk) 
&\to 
	\frac{\bqt^2\, K_c(\sk)}{12\, (1+Q^2)} \, \log r +\order{r^0}\,.
\end{split}
\end{equation}	
This follows from the fact that these fields are dual to long-lived boundary modes. Given this behaviour, one can check that the radial integrals appearing in~\eqref{eq:IFFPintegrals} encounter no UV divergences from the large $r$ region. The integrals can be therefore evaluated, at the very least numerically, quite efficiently to any given order in the gradient expansion. We give results for a sampling of terms below.

There is, however, one subtlety that has to be dealt with in the evaluation of the correlation functions. The radial derivative operators acting on the boundary-bulk Green's functions $\Gin{}(\sk)$, or its time reversed counterpart $\Gin{}(\bar{\sk})$, are innocuous. But when the derivative hits the radial Boltzmann factor, $e^{-\beta\omega\ctor}$ it brings down a factor of $1/f$ owing to~\eqref{eq:mockT}. Since the Green's functions and their derivatives are regular at the horizon (by construction), it then naively implies that the integrals~\eqref{eq:IFFPintegrals} are IR divergent. Fortunately, the grSK contour integral prescription takes this into account, and gives a clear way to regulate this potential divergence. Factors of $1/f$ (or powers thereof) lead to singularities at the branch point, and in addition to the discontinuity across the cut, one has an additional localized contribution. As demonstrated in~\cite{Loganayagam:2022teq} (see also~\cite{Chakrabarty:2019aeu} for an earlier analysis) divergence from these two factors cancels, leading to a finite answer for the correlation function. 
Likewise, there are no subtleties associated with the zeros of the function $\Lk$, which appears in the scalar sector. While $\Vd$ is singular at these points ($\Zd$ is regular), the function $\MV$ is regular there. There are no singularities encountered along the grSK contour for any (even complex) momenta. We illustrate this below while presenting our results.

\begin{figure}[t]\label{Fig:VVVCoeff2}
\subfloat{
\begin{minipage}[h!]{0.4\textwidth}
\begin{tikzpicture}
  \node (img)  {\includegraphics[scale=0.75]{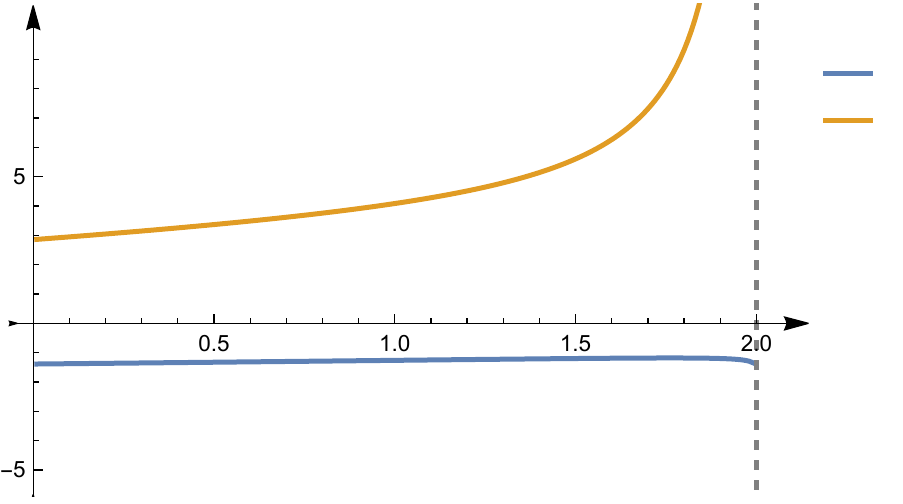}};
  \node[below=of img, node distance=0cm,font=\small,yshift=2.5cm,xshift=2.8cm] {$Q^2$};
  \node[below=of img, node distance=0cm, anchor=center,font=\small,yshift=4.4cm,xshift=3.55cm] {$\mathfrak{a}_1$};
  \node[below=of img, node distance=0cm, anchor=center,font=\small,yshift=4.0cm,xshift=3.55cm] {$\mathfrak{a}_2$};
 \end{tikzpicture}
 \end{minipage}}
\hspace{1.5cm}
 \subfloat{
\begin{minipage}[h!]{0.4\textwidth}
\begin{tikzpicture}
  \node (img)  {\includegraphics[scale=0.75]{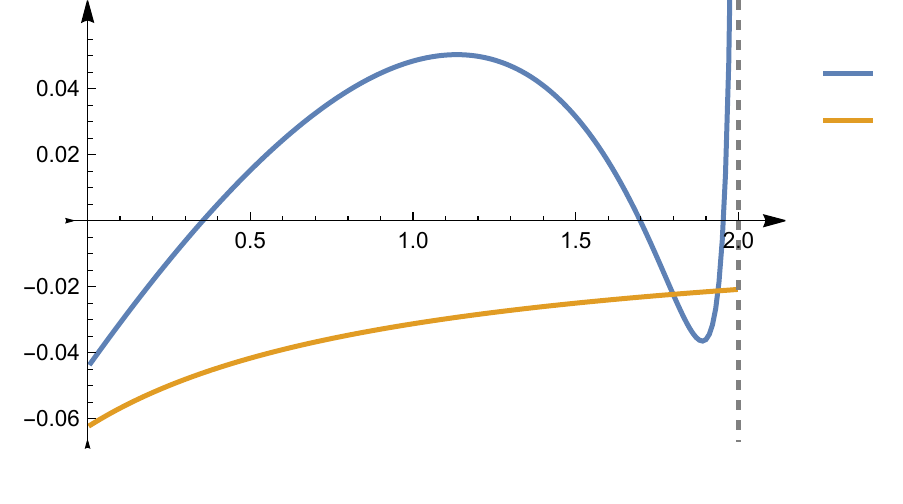}};
  \node[below=of img, node distance=0cm,font=\small,yshift=3.3cm,xshift=2.8cm] {$Q^2$};
  \node[below=of img, node distance=0cm, anchor=center,font=\small,yshift=4.4cm,xshift=3.55cm] {$\mathfrak{a}_3$};
  \node[below=of img, node distance=0cm, anchor=center,font=\small,yshift=4.0cm,xshift=3.55cm] {$\mathfrak{a}_4$};
 \end{tikzpicture}
 \end{minipage}}
 \caption{Plots of the coefficients parameterizing $I_{_\text{PPF}}^{\eta_1\eta_2\eta_3}$ as defined in~\eqref{eq:PPF3Ps} appearing in the gradient expansion of the $\POp_{_\text{P}}^{\eta_1} \POp_{_\text{P}}^{\eta_2} \POp_{_\text{F}}^{\eta_3}$ cubic term in the anomalous effective action.   }
 \label{fig:VVVCoeff2}
 \end{figure}
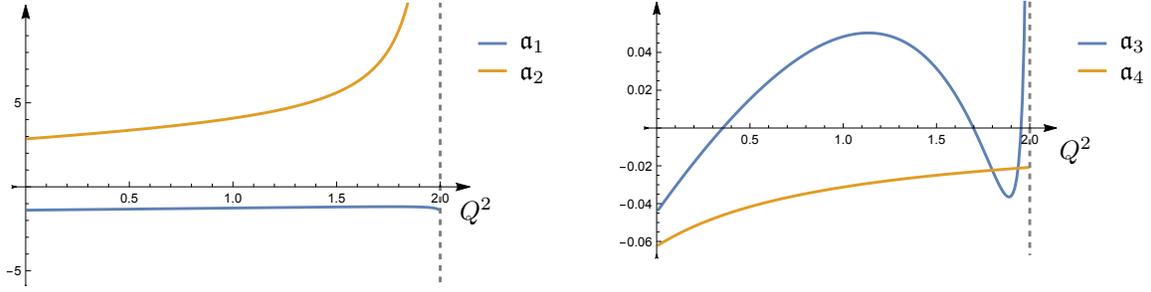

\paragraph{The 3-point shear vertices:} Since the data for the field $\MX_\eta$ is given in detail in~\cref{sec:vectors}, we start with the contributions~\eqref{eq:PPPterms}. While we can factor out the polarization dependence using the tensor structure $\mathbb{N}^{\eta_1\eta_2\eta_3}$ introduced in~\eqref{eq:CSvvv}, we still have to account for the asymmetry arising from one of the operators having a derivative acting on it. Factoring this in, one has the contribution to the Wilsonian influence functional to be
\begin{equation}
\begin{split}
S_{_\text{WIF}}^{(3)}
&\supset
	\mathbb{N}^{\eta_1\eta_2\eta_3} \bigg[\int_{\sk_1,\sk_2,\sk_3}
	I_{_\text{FFP}}^{\eta_1\eta_2\eta_3} \, \Kin{\text{shear}}(\sk_1) \, \Kin{\text{shear}}(\sk_2)\, \Kin{\text{shear}}(\bar{\sk}_3)
		\POp_{_\text{F}}^{\eta_1}(\sk_1) \POp_{_\text{F}}^{\eta_2}(\sk_2) \POp_{_\text{P}}^{\eta_3}(\sk_3)\\
&\qquad 
	+ \int_{\sk_1,\sk_2,\sk_3}
	I_{_\text{PPF}}^{\eta_1\eta_2\eta_3} \, \Kin{\text{shear}}(\bar{\sk}_1) \, \Kin{\text{shear}}(\bar{\sk}_2)\, \Kin{\text{shear}}(\sk_3) \, \POp_{_\text{P}}^{\eta_1}(\sk_1) \POp_{_\text{P}}^{\eta_2}(\sk_2) \POp_{_\text{F}}^{\eta_3}(\sk_3) \bigg] \,.
\end{split}
\end{equation}	
The 3-point  functions are themselves given by
\begin{equation}\label{eq:3shearI}
\begin{split}
	I_{_\text{FFP}}^{\eta_1\eta_2\eta_3} \,
&
=
 \left(1-e^{\beta\omega_3}\right)
	\int_{r_+}^{r_c}\, dr\,
	\, \frac{\Gin{\MX_{\eta_1}}(\ctor, \sk_1)}{r^2} \, \frac{\Gin{\MX_{\eta_2}}(\ctor, \sk_2)}{r^2} \, \dv{r}( e^{-\beta\omega_3\ctor} \frac{\Gin{\MX_{\eta_3}}(\ctor, \bar{\sk}_3)}{r^2}) ,
\\
I_{_\text{PPF}}^{\eta_1\eta_2\eta_3} 
&
=
	 \left(1-e^{-\beta\omega_3}\right)
	\int_{r_+}^{r_c}\, dr\, e^{\beta\omega_3\ctor} 
	\, \frac{\Gin{\MX_{\eta_1}}(\ctor,\bar{\sk}_1)}{r^2} \, \frac{\Gin{\MX_{\eta_2}}(\ctor,\bar{\sk}_2)}{r^2} \, \dv{r}(\frac{\Gin{\MX_{\eta_3}}(\ctor, \sk_3)}{r^2}) \,.
\end{split}
\end{equation}
As noted the integrals can be evaluated in a straightforward manner. For simplicity, focusing on the second integral, which has no localized contribution from the horizon, we find
\begin{equation}\label{eq:PPF3Ps}
\begin{split}
	I_{_\text{PPF}}^{\eta_1\eta_2\eta_3} 
&=
	-\frac{\nB(\omega_3)}{\nB(\omega_3)+1}
	\left(2 + i\, \mathfrak{a}_1\,\bwt_3 +\mathfrak{a}_2\, \bwt_3^2 - \mathfrak{a}_3 (\bwt_1^2+\bwt_2^2)
	+ \mathfrak{a}_4 \, (\bqt_1^2+\bqt_2^2) + \cdots \right), 
\end{split}
\end{equation}
In writing the result, we have judiciously applied energy and momentum conservation to simplify terms.  The coefficients $\mathfrak{a}_i$ are plotted in~\cref{fig:VVVCoeff2}.

\begin{figure}[t]
\subfloat{
\begin{minipage}[h!]{0.5\textwidth}
\begin{tikzpicture}
  \node (img)  {\includegraphics[scale=0.75]{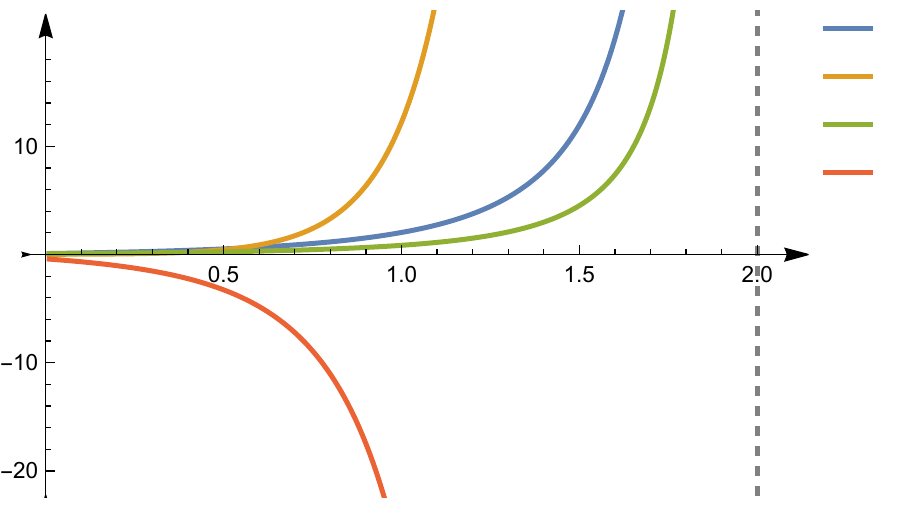}};
  \node[below=of img, node distance=0cm,font=\small,yshift=3.2cm,xshift=2.9cm] {$Q^2$};
  \node[below=of img, node distance=0cm, anchor=center,font=\tiny,yshift=4.8cm,xshift=3.7cm] {$\mathfrak{g}_1^{\QOp}$};
  \node[below=of img, node distance=0cm, anchor=center,font=\tiny,yshift=4.4cm,xshift=3.7cm] {$\mathfrak{g}_2^{\QOp}$};
  \node[below=of img, node distance=0cm, anchor=center,font=\tiny,yshift=4.0cm,xshift=3.7cm] {$\mathfrak{g}_3^{\QOp}$};
  \node[below=of img, node distance=0cm, anchor=center,font=\tiny,yshift=3.7cm,xshift=3.7cm] {$\mathfrak{g}_4^{\QOp}$};
 \end{tikzpicture}
 \end{minipage}}
 \subfloat{
\begin{minipage}[h!]{0.5\textwidth}
\hspace{0.3cm}
\begin{tikzpicture}
  \node (img)  {\includegraphics[scale=0.75]{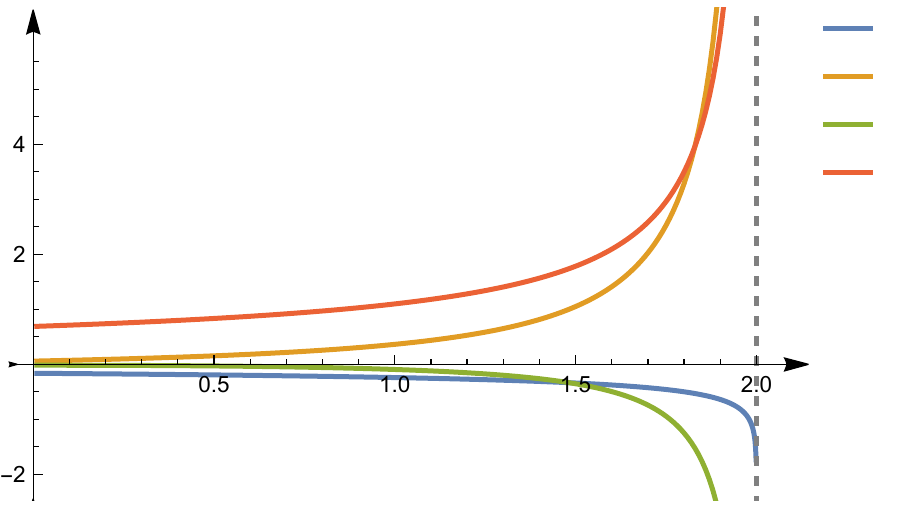}};
  \node[below=of img, node distance=0cm,font=\small,yshift=2.3cm,xshift=2.9cm] {$Q^2$};
 \node[below=of img, node distance=0cm, anchor=center,font=\tiny,yshift=4.8cm,xshift=3.6cm] {$\mathfrak{g}_1^\EOp$};
  \node[below=of img, node distance=0cm, anchor=center,font=\tiny,yshift=4.4cm,xshift=3.6cm] {$\mathfrak{g}_2^\EOp$};
  \node[below=of img, node distance=0cm, anchor=center,font=\tiny,yshift=4.0cm,xshift=3.6cm] {$\mathfrak{g}_3^\EOp$};
  \node[below=of img, node distance=0cm, anchor=center,font=\tiny,yshift=3.6cm,xshift=3.6cm] {$\mathfrak{g}_4^\EOp$};
 \end{tikzpicture}
 \end{minipage}}
 \caption{Plots of the coefficients parameterizing $I_{_\text{FFP}}^{\eta}$ as defined in~\eqref{eq:FFP1P} appearing in the gradient expansion of the $\mathcal{S}_{_\text{F}} \,\mathcal{S} _{_\text{F}} \POp_{_\text{P}}^{\eta}$ cubic term in the anomalous effective action. }
 \label{fig:SSVCoeff2}
 \end{figure}
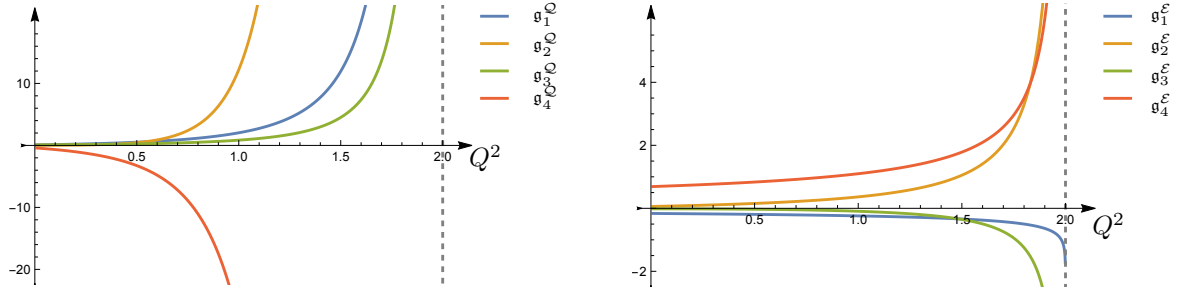

\paragraph{Vertices with a single shear mode:} The vertex~\eqref{eq:CSvss}, which is the next simplest to evaluate, contributes terms with a single chiral shear mode. Continuing to use $\mathcal{S}$ to denote either $\QOp$ or $\EOp$, we have the contribution to the influence functional as recorded in~\eqref{eq:PSSterms}. We will illustrate the general features with a particular term,  viz., 
\begin{equation}
\begin{split}
S_{_\text{WIF}}^{(3)}
\supset
	\int_{\sk_1,\sk_2,\sk_3} 
	I_{_\text{FFP}}^\eta\, \Kin{\mathcal{S} }(\sk_1)\, \Kin{\mathcal{S} }(\sk_2)\,  \Kin{\text{shear}}(\bar{\sk}_3)\, \mathcal{S}_{_\text{F}}(\sk_1)\, \mathcal{S}_{_\text{F}}(\sk_2)\, \POp_{_\text{P}}^\eta(\sk_3)\,. 
\end{split}
\end{equation} 
The 3-point functions contributing to these terms is can be obtained from the master formula
\begin{equation}
\begin{split}
I_{_\text{FFP}}^\eta
&= 
	\mathbb{N}^{\eta} 
	\left( 1-e^{\beta\omega_{3} } \right)
	\int_{r_+}^\infty\, dr \, \frac{\Gin{\MV}(\sk_1)}{r^2}\,\dv{\Gin{\MV}(\sk_2)}{r}\, e^{-\beta\omega_3\ctor} \frac{\Gin{\MX_{\eta}}(\bar{\sk}_3)}{r^2}\,,
\\
\mathbb{N}^{\eta} 
&=
		\frac{8\, \kcs \,\RQ^2}{3\,\mu}\,\omega_1\,\bqt_3^2 \; 
		    \VSSa^{\eta}(\sk_1,\sk_2,\sk_3)\,.
\end{split}
\end{equation}	
Since $\MV$ is a linear combination of $\Vd$ and $\Zd$, we can immediately extract the three independent contributions $\QOp_{_\text{P}}^\eta(\sk_1)\,\QOp_{_\text{P}}^\eta(\sk_2) \,\POp_{_\text{P}}^\eta(\sk_3)$,
$\QOp_{_\text{P}}^\eta(\sk_1)\,\EOp_{_\text{P}}^\eta(\sk_2) \,\POp_{_\text{P}}^\eta(\sk_3)$, and $\EOp_{_\text{P}}^\eta(\sk_1)\,\EOp_{_\text{P}}^\eta(\sk_2) \,\POp_{_\text{P}}^\eta(\sk_3)$ to $S_{_\text{WIF}}^{(3)}$. The final result can be written as 
\begin{equation}\label{eq:FFP1P}
\begin{split}
I_{_\text{FFP}}^\eta[\QOp ,\QOp ,\POp] 
&= 
	\frac{\mathbb{N}^{\eta}}{4}\,  
	\left( 1-e^{\beta\omega_{3} } \right) \left[ 
	 -i\, \mathfrak{g}_1^{\QOp}\,\bwt_2  - \mathfrak{g}_2^{\QOp}\, \bwt_2^2 +\mathfrak{g}_3 ^{\QOp}\, \bqt_2^2
	-\mathfrak{g}_4^{\QOp}\, \bwt_1\,\bwt_2 \right] ,\\
I_{_\text{FFP}}^\eta[\EOp ,\EOp ,\POp] 
&= 
	\frac{\mathbb{N}^{\eta}}{9\,\bRQ^2}\, 
	\left( 1-e^{\beta\omega_{3} } \right) \left[ 
	 -i\, \mathfrak{g}_1^\EOp\,\bwt_2 
	- \mathfrak{g}_2^\EOp\, \bwt_2^2 
	+ \mathfrak{g}_3^\EOp\, \bqt_2^2 
	-\mathfrak{g}_4^\EOp\, \bwt_1\,\bwt_2\right] , \\
I_{_\text{FFP}}^\eta[\QOp ,\EOp ,\POp] 
&= 
	-\frac{ \mathbb{N}^{\eta}}{6\,\bRQ}\, 
	\left( 1-e^{\beta\omega_{3} } \right) \left[ 
	-i\, (\mathfrak{g}_1^{\QOp}+\mathfrak{g}_1^\EOp)\,\bwt_2  
	- (\mathfrak{g}_2^{\QOp}+\mathfrak{g}_2^\EOp)\, \bwt_2^2
	\right. \\
&\left. \qquad \qquad \qquad \qquad \quad
	+ (\mathfrak{g}_3^{\QOp}+\mathfrak{g}_3^\EOp)\, \bqt_2^2 
	-(\mathfrak{g}_4^{\QOp}+\mathfrak{g}_4^\EOp)\, \bwt_1\,\bwt_2
 	\right] \,.
\end{split}
\end{equation}
We focused on this choice to avoid computing the localized contribution from the outgoing propagator $\Gout{\MV}$ and have again used energy and momentum conservation to simplify terms.  The coefficients $\mathfrak{g}_i^{\QOp}$ and $\mathfrak{g}_i^\EOp$ are plotted in~\cref{fig:SSVCoeff2}. 

\paragraph{Vertices with two shear modes:} Finally, let us look at the contribution from the vertex~\eqref{eq:CSvvs}, which has the most intricate structure. Here it is easy to see that one will have a localized contribution at the horizon. We will again focus on a particular assignment of advanced retarded labels to illustrate the physics. Consider
\begin{equation}
\begin{split}
S_{_\text{WIF}}^{(3)}
&\supset
	\int_{\sk_1,\sk_2,\sk_3}
	I_{_\text{FFP}}^{\eta_1\eta_2} \, \Kin{\mathcal{S}}(\sk_1) \, \Kin{\text{shear}}(\sk_2)\, \Kin{\text{shear}}(\bar{\sk}_3)\, 
		\mathcal{S}_{_\text{F}} (\sk_1) \POp_{_\text{F}}^{\eta_1}(\sk_2) \POp_{_\text{P}}^{\eta_2}(\sk_3)\,,
\end{split}	
\end{equation}
where $\mathcal{S} \in \{\QOp,\EOp\}$. The 3-point function of interest can be decomposed into three independent tensor structures
\begin{equation}\label{eq:IFFP2v}
\begin{split}
I_{_\text{FFP}}^{\eta_1\eta_2} 
&= 
	\left( 1-e^{-\beta\omega_3} \right) \left( \mathbb{N}^{\eta_1\eta_2}_1\, J_1 + \mathbb{N}^{\eta_1\eta_2}_2\, J_2 +\mathbb{N}^{\eta_1\eta_2}_3\, J_3 \right)  .
\end{split}
\end{equation}
The form factors involved above are
\begin{equation}
\begin{split}
\mathbb{N}^{\eta_1\eta_2}_1
&=
	-\frac{2\,\RQ^4}{3\,\mu^2} \kcs\, \bqt_2^2\,\bqt_3^2 \left( \VVSa^{\eta_1\eta_2}(\sk_1,\sk_2,\sk_3) + \VVSa^{\eta_2\eta_1}(\sk_1,\sk_3,\sk_2) \right) , \\
\mathbb{N}^{\eta_1\eta_2}_2
&=
	-\frac{2\,\RQ^4}{3\,\mu^2} \kcs\, \bqt_2^2\,\bqt_3^2 \left( 2\, \VVSa^{\eta_1\eta_2}(\sk_1,\sk_2,\sk_3) - \VVSa^{\eta_2\eta_1}(\sk_1,\sk_3,\sk_2) \right) ,\\
\mathbb{N}^{\eta_1\eta_2}_3
&=
	\frac{2\,\RQ^4}{3\,\mu^2} \kcs\, \bqt_2^2\,\bqt_3^2 \left( 2\, \VVSa^{\eta_1\eta_2}(\sk_1,\sk_2,\sk_3) -2\, \VVSa^{\eta_2\eta_1}(\sk_1,\sk_3,\sk_2) \right) .
\end{split}
\end{equation}
The contributions $J_i$ are radial integrals involving the boundary-bulk propagators, and are explicitly given to  be
\begin{equation}
\begin{split}
J_1 
&=
	\int_{r_+}^\infty\, dr\, \frac{e^{-\beta\omega_3\ctor}}{r^3}\, \Gin{\MX_{\eta_1}}(\sk_2)\, \Gin{\MX_{\eta_2}}(\bar{\sk}_3)\left[ \frac{1}{r}\,\dv{r}(r f\,\dv{r}) -\frac{i\omega_2}{r}\,\dv{r}\right] \frac{\Gin{\MV}(\sk_1)}{r}, \\
J_2 
&=
	\int_{r_+}^\infty\, dr\, \frac{e^{-\beta\omega_3\ctor}}{r^3}\,  \Gin{\MX_{\eta_2}}(\bar{\sk}_3)
	\left[\Dz_+ \Gin{\MV}(\sk_1)\,\dv{r}(\frac{\Gin{\MX_{\eta_1}}(\sk_2)}{r^2}) -i\,\omega_2\, \dv{\Gin{\MV}(\sk_1)}{r}\, \frac{\Gin{\MX_{\eta_1}}(\sk_2)}{r^2}
	\right] ,\\
J_3 
&=
	\int_{r_+}^\infty\, dr\, \frac{e^{-\beta\omega_3\ctor}}{r^3}\,  \Gin{\MX_{\eta_1}}(\sk_2)
	\left[\Dz_+ \Gin{\MV}(\sk_1)\,\dv{r}(\frac{\Gin{\MX_{\eta_2}}(\bar{\sk}_3)}{r^2}) +3i\,\omega_2\, \dv{\Gin{\MV}(\sk_1)}{r}\, \frac{\Gin{\MX_{\eta_2}}(\bar{\sk}_3)}{r^2}
	\right] \\
& \qquad 
	+\frac{2\omega_1\omega_3}{1-e^{-\beta\omega_3}} \oint_{D_\epsilon}\, \frac{e^{-\beta\omega_3\ctor}}{r^7 f}\, \Gin{\MV}(\sk_1)\, \Gin{\MX_{\eta_1}}(\sk_2)\, \Gin{\MX_{\eta_2}}(\bar{\sk}_3) \,.
\end{split}
\end{equation}
The integrals contributing to  $J_1$ and $J_2$ are similar to  the ones we have hitherto  computed. However, $J_3$ has a localized contribution from the radial derivative acting on the outgoing propagator of $\MX_{\eta_2}(\sk_3)$, which is evaluated as a contour integral over an infinitesimal disc $D_\epsilon$ centered at $r_+$. Notice that have already factored out $1-e^{-\beta\omega_3}$, the factor that arises from computing the discontinuity across the cut, in writing~\eqref{eq:IFFP2v}. We are thus dividing the localized contribution to $J_3$ by this factor (a similar factor arises from the evaluation of the $D_\epsilon$ contour integral, canceling this denominator at the end).

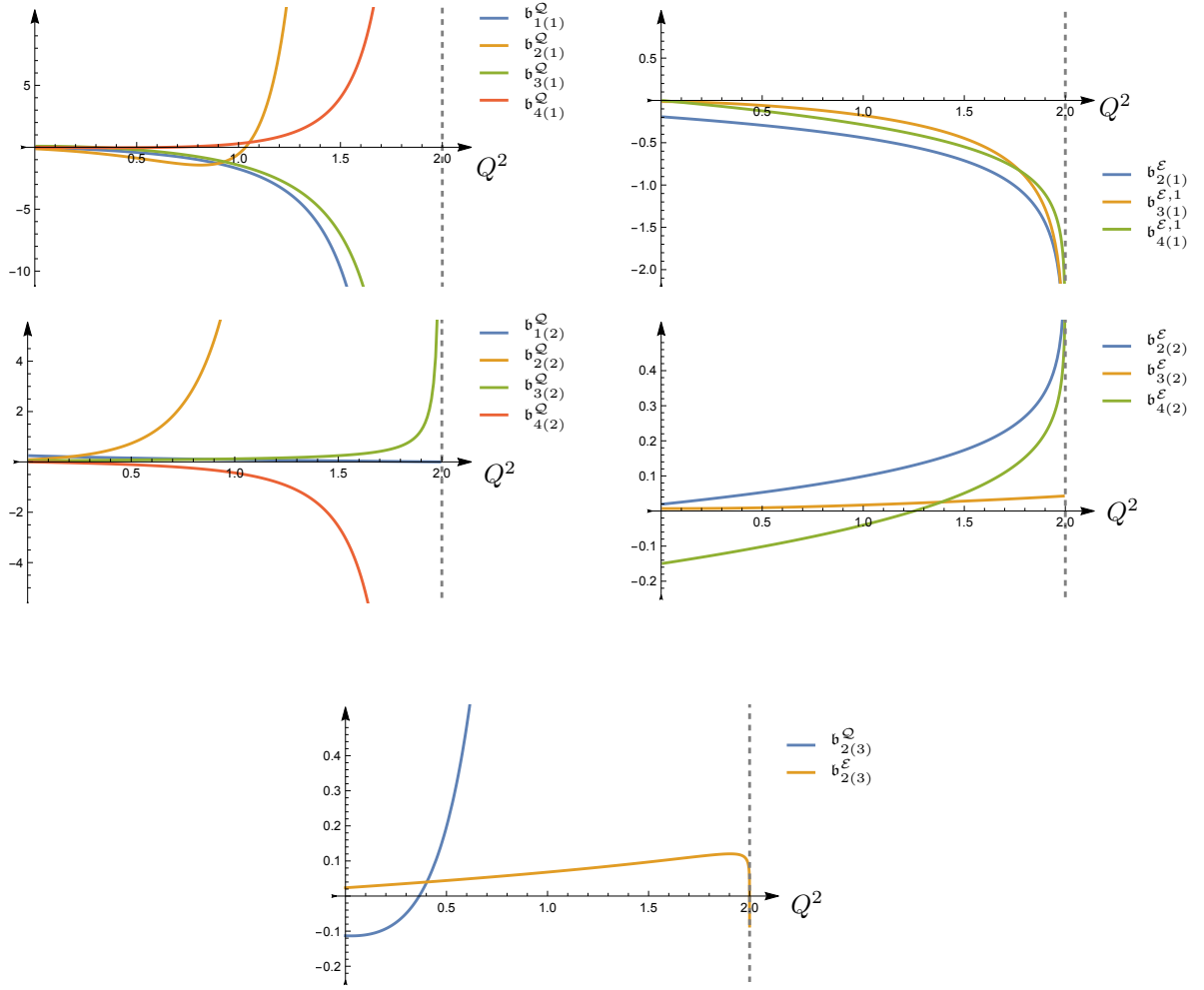
\begin{figure}[t]
\subfloat{
\begin{minipage}[h!]{0.5\textwidth}
\begin{tikzpicture}
  \node (img)  {\includegraphics[scale=0.75]{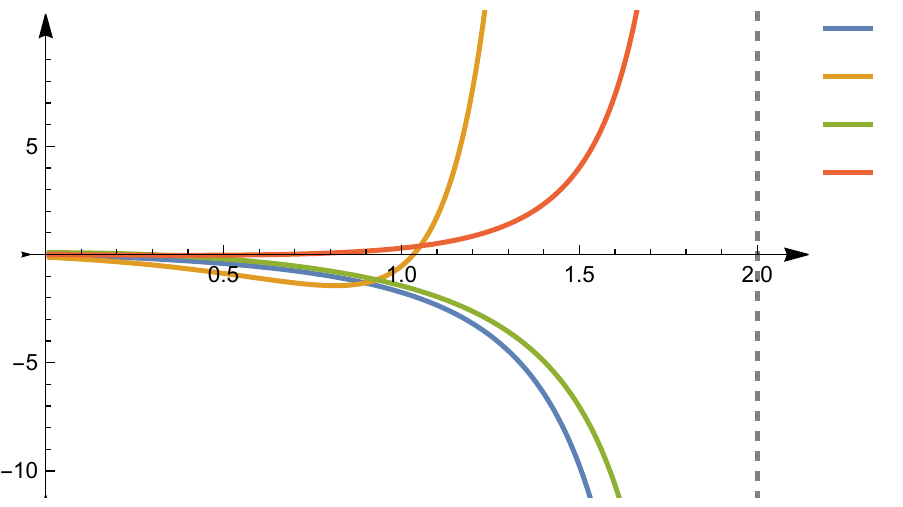}};
  \node[below=of img, node distance=0cm,font=\small,yshift=3.1cm,xshift=3.0cm] {$Q^2$};
  \node[below=of img, node distance=0cm, anchor=center,font=\tiny,yshift=4.8cm,xshift=3.7cm] {$\mathfrak{b}^\QOp_{1(1)}$};
  \node[below=of img, node distance=0cm, anchor=center,font=\tiny,yshift=4.4cm,xshift=3.7cm] {$\mathfrak{b}^\QOp_{2(1)}$};
  \node[below=of img, node distance=0cm, anchor=center,font=\tiny,yshift=4.0cm,xshift=3.7cm] {$\mathfrak{b}^\QOp_{3(1)}$};
  \node[below=of img, node distance=0cm, anchor=center,font=\tiny,yshift=3.6cm,xshift=3.7cm] {$\mathfrak{b}^\QOp_{4(1)}$};
 \end{tikzpicture}
 \end{minipage}}
 \subfloat{
\begin{minipage}[h!]{0.5\textwidth}
\hspace{0.3cm}
\begin{tikzpicture}
  \node (img)  {\includegraphics[scale=0.75]{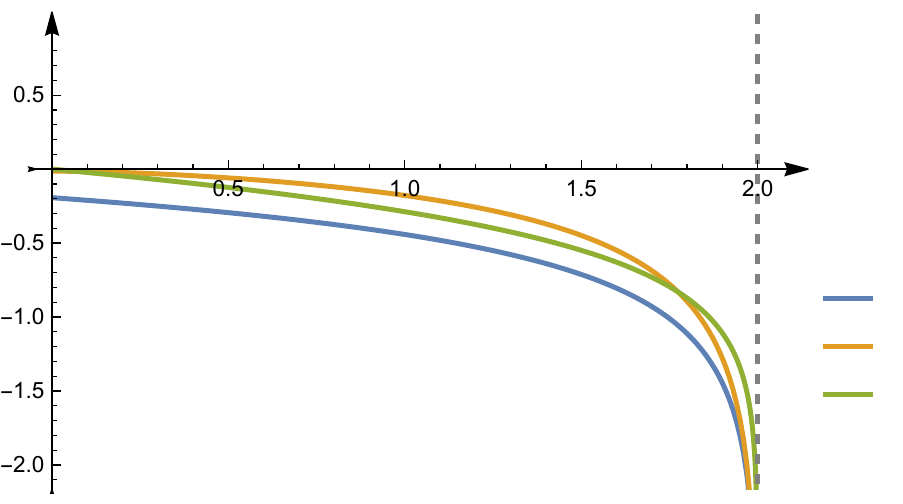}};
  \node[below=of img, node distance=0cm,font=\small,yshift=3.9cm,xshift=3.0cm] {$Q^2$};
  \node[below=of img, node distance=0cm, anchor=center,font=\tiny,yshift=2.7cm,xshift=3.7cm] {$\mathfrak{b}^{\EOp}_{2(1)}$};
  \node[below=of img, node distance=0cm, anchor=center,font=\tiny,yshift=2.3cm,xshift=3.7cm] {$\mathfrak{b}^{\EOp,1}_{3(1)}$};
  \node[below=of img, node distance=0cm, anchor=center,font=\tiny,yshift=1.9cm,xshift=3.7cm] {$\mathfrak{b}^{\EOp,1}_{4(1)}$};
 \end{tikzpicture}
 \end{minipage}}\\
 \subfloat{
\begin{minipage}[h!]{0.5\textwidth}
\begin{tikzpicture}
  \node (img)  {\includegraphics[scale=0.75]{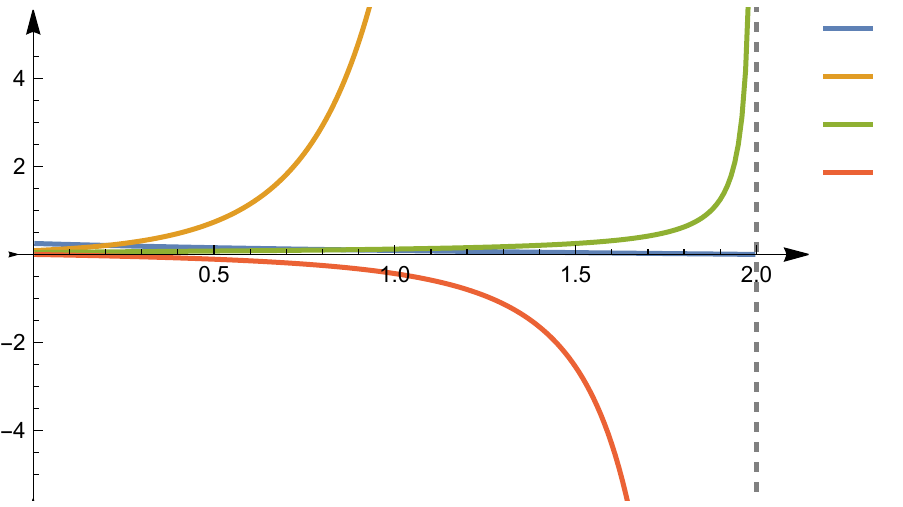}};
  \node[below=of img, node distance=0cm,font=\small,yshift=3.2cm,xshift=3.1cm] {$Q^2$};
  \node[below=of img, node distance=0cm, anchor=center,font=\tiny,yshift=4.85cm,xshift=3.7cm] {$\mathfrak{b}^\QOp_{1(2)}$};
  \node[below=of img, node distance=0cm, anchor=center,font=\tiny,yshift=4.45cm,xshift=3.7cm] {$\mathfrak{b}^\QOp_{2(2)}$};
  \node[below=of img, node distance=0cm, anchor=center,font=\tiny,yshift=4.05cm,xshift=3.7cm] {$\mathfrak{b}^\QOp_{3(2)}$};
  \node[below=of img, node distance=0cm, anchor=center,font=\tiny,yshift=3.65cm,xshift=3.7cm] {$\mathfrak{b}^\QOp_{4(2)}$};
 \end{tikzpicture}
 \end{minipage}}
 \subfloat{
\begin{minipage}[h!]{0.5\textwidth}
\hspace{0.3cm}
\begin{tikzpicture}
  \node (img)  {\includegraphics[scale=0.75]{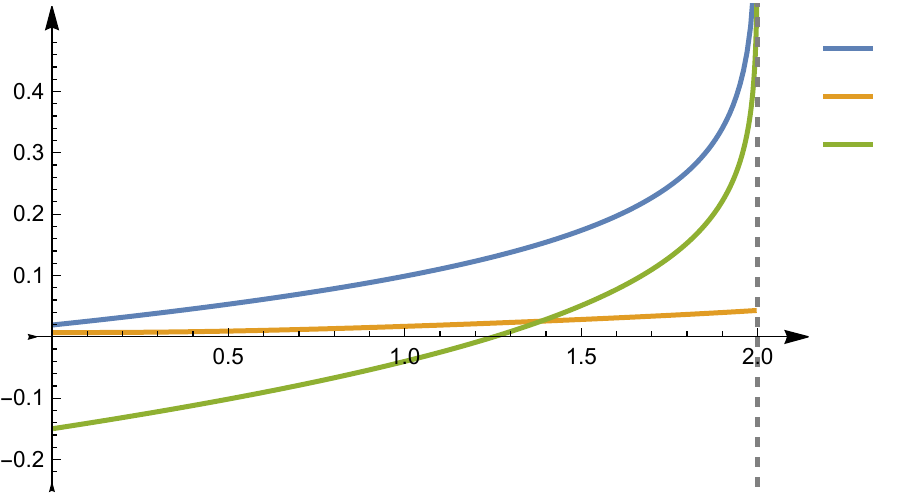}};
  \node[below=of img, node distance=0cm,font=\small,yshift=2.7cm,xshift=3.1cm] {$Q^2$};
  \node[below=of img, node distance=0cm, anchor=center,font=\tiny,yshift=4.65cm,xshift=3.7cm] {$\mathfrak{b}^\EOp_{2(2)}$};
  \node[below=of img, node distance=0cm, anchor=center,font=\tiny,yshift=4.25cm,xshift=3.7cm] {$\mathfrak{b}^\EOp_{3(2)}$};
  \node[below=of img, node distance=0cm, anchor=center,font=\tiny,yshift=3.85cm,xshift=3.7cm] {$\mathfrak{b}^\EOp_{4(2)}$};
 \end{tikzpicture}
 \end{minipage}}\\
\subfloat{
\begin{minipage}[h!]{0.5\textwidth}
\vspace{1cm}
\hspace{4cm}
\begin{tikzpicture}
  \node (img)  {\includegraphics[scale=0.75]{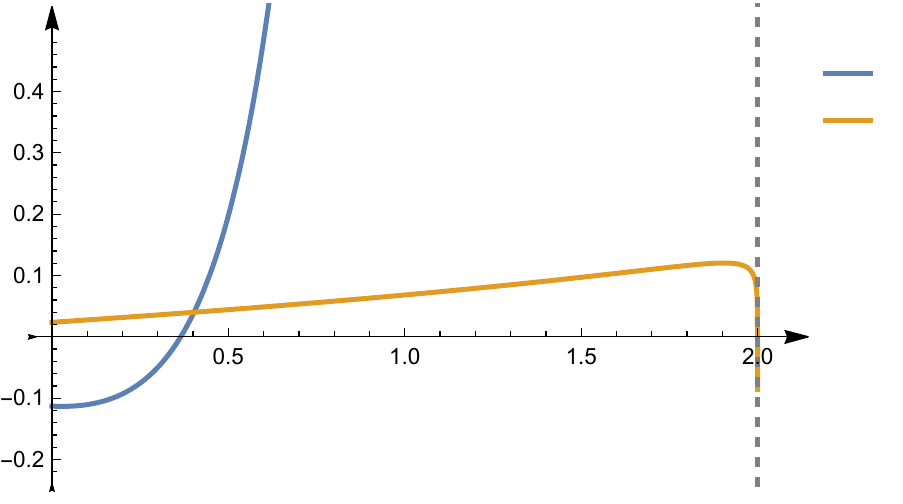}};
  \node[below=of img, node distance=0cm,font=\small,yshift=2.6cm,xshift=3.1cm] {$Q^2$};
  \node[below=of img, node distance=0cm, anchor=center,font=\tiny,yshift=4.45cm,xshift=3.7cm] {$\mathfrak{b}^\QOp_{2(3)}$};
  \node[below=of img, node distance=0cm, anchor=center,font=\tiny,yshift=4.05cm,xshift=3.7cm] {$\mathfrak{b}^\EOp_{2(3)}$};
 \end{tikzpicture}
 \end{minipage}}
 \caption{Plots of the coefficients parameterizing $I_{_\text{FFP}}^{\eta_1\eta_2}$ as defined in~\eqref{eq:FFP2P} appearing in the gradient expansion of the $\mathcal{S} {_\text{F}} \,\mathcal{P}^{\eta_1}_{_\text{F}} \POp^{\eta_2}_{_\text{P}}$ cubic term in the anomalous effective action. }
 \label{fig:SVVCoeff2}
 \end{figure}

In any event all of these integrals can be evaluated as before and lead to 
\begin{equation}\label{eq:FFP2P}
\begin{split}
I_{_\text{FFP}}^{\eta_1\eta_2}[\QOp,\POp_{\eta_1}, \POp_{\eta_2}] 
&= 
	\left( 1-e^{-\beta\omega_3} \right) \bigg(
	\sum_{j=1}^3\, \mathbb{N}^{\eta_1\eta_2}_j\, \left[-i\, \mathfrak{b}_{1(j)}^\QOp \, \bwt_1-
	\mathfrak{b}_{2(j)}^\QOp\, \bwt_1^2 
	+\mathfrak{b}_{3(j)}^\QOp\, \bqt_1^2 
	- \mathfrak{b}_{4(j)}^\QOp\,\bwt_1\,\bwt_2   \right]  \\
& \qquad \qquad \qquad \qquad 
	-2\,  \bwt_1\,\bwt_3\, \mathfrak{b}_\text{loc} \bigg),\\
I_{_\text{FFP}}^{\eta_1\eta_2}[\EOp,\POp_{\eta_1}, \POp_{\eta_2}] 
&= 
	\frac{2}{3\,\bRQ} \left( 1-e^{-\beta\omega_3} \right) 
	\bigg(\sum_{j=1}^3\, \mathbb{N}^{\eta_1\eta_2}_j\, \left[-i\, \mathfrak{b}_{1(j)}^\EOp \, \bwt_1-
	\mathfrak{b}_{2(j)}^\EOp\, \bwt_1^2 
	+\mathfrak{b}_{3(j)}^\EOp\, \bqt_1^2 
	- \mathfrak{b}_{4(j)}^\EOp\,\bwt_1\,\bwt_2   \right] \\
& \qquad \qquad \qquad \qquad 
	+2\,  \bwt_1\,\bwt_3\, \mathfrak{b}_\text{loc} \bigg)
\end{split}
\end{equation}
Explicit evaluation of these coefficients leads to relations
\begin{equation}
\mathfrak{b}_{j(2)}^\QOp = \mathfrak{b}_{j(3)}^\QOp \,,
\qquad 
\mathfrak{b}_{j(2)}^\EOp = \mathfrak{b}_{j(3)}^\EOp \,,
\qquad 
\text{for} \; j =1,3,4\,.
\end{equation}	
In addition, the coefficients of the linear term in frequency turn out to be quite simple
\begin{equation}
\begin{split}
\mathfrak{b}_{1(1)}^\QOp &= -\frac{(8-Q^2)\,Q^2}{4(2-Q^2)^2}\,, \qquad
\mathfrak{b}_{1(1)}^\EOp = \frac{1}{4}	
\,, \qquad
\mathfrak{b}_{1(2)}^\EOp = -\frac{1}{8}	\,.
\end{split}
\end{equation}
Likewise, the localized contribution from $J_3$ is also analytic up to the order we have computed, and is given to be 
\begin{equation}
\mathfrak{b}_\text{loc} =
\frac{1}{Q^2\,(2-Q^2)}
	\left[\left(\frac{49}{40}- \frac{\log 2}{2}\right)Q^2+\frac{1}{2} \,\log(2-Q^2)-\frac{(1+Q^2)}{\sqrt{1+4\,Q^2}}\, 
	\coth^{-1}\left(\frac{3}{\sqrt{1+4\,Q^2}}\right) \right]	
\end{equation}
The remaining coefficients are straightforward to evaluate numerically, and are plotted in~\cref{fig:SVVCoeff2}.

\section{Discussion}\label{sec:discuss}

The main thrust of our discussion was to derive the hydrodynamic effective action, using the gravitational Schwinger-Keldysh (grSK) formalism, for a charged fluid with an anomalous  (abelian) global symmetry. We focused on deriving an action,  with the dynamical fields parameterizing small amplitude fluctuations in the vicinity of the homogeneous global equilibrium solution. At the quadratic order the anomaly is confined to momentum diffusion. One finds chiral shear waves with anomaly dependent dispersion relations, starting at the cubic order in gradients (at quadratic order the shear dispersion is fixed by shear viscosity). The dynamics of charge diffusion and energy transport is unmodified by the anomaly at the Gaussian order. In particular, there is no contribution from the anomaly to the charge diffusion and sound mode dispersion relations (at any order).

At the interacting level, however,  all the hydrodynamic fields are affected by the anomaly. The Chern-Simons term in the holographic description, which captures the 't Hooft anomaly induces non-trivial cubic couplings between the different hydrodynamic modes. These couplings are qualitatively different from other cubic couplings present in charged plasmas. For one, they are parity-odd, and for another they are uniquely fixed by the structure of the anomaly (which will become transparent when we discuss gauge invariance below). Grouping the charge diffusion and energy transport into a single scalar field ($\MV$ as in the text),  in the hydrodynamic effective field theory we find the following cubic terms: (i) coupling between three shear modes, or (ii) two shear modes coupling a scalar mode, and (iii) coupling between a single shear mode to two scalar modes. As the fields also  carry advanced/retarded labels the result set of non-Gaussian terms is quite large.
We have given the general rules for computing the correlation functions, which determine the cubic anomaly induced vertices the effective field theory, and have computed exemplars in each of the three aforementioned cases.

From a physical perspective, the interactions between the diffusive (shear and charge) and sound modes are interesting in their own right. If we were to consider the analog of a scattering problem, e.g., the correlation of four shear modes in a fluid, we learn that there is an intermediate channel with a propagating sound mode, whose (parity-odd) residue is fixed by the anomaly. This factorization can be explored  along the lines explained in~\cite{Loganayagam:2022zmq}. The general scheme outlined here provides the necessary ingredients for setting up the appropriate crossing equations. 

There are several important issues that we have not directly addressed in the analysis. One the primary motivations for the analysis was to ascertain the structure of the anomaly constrained contribution to the effective action. This in part, was to verify the structure of the effective action argued for in~\cite{Haehl:2013hoa}. In writing an effective action, one has to confront the usual issue with 't Hooft anomalies: there is no gauge invariant generating functional of the background sources in the same dimension as the original quantum system. However, using the inflow mechanism, one can construct a gauge invariant action. One uses the fact that anomaly of the current associated with $A^\text{b}_\mu$ can be canceled by a coupling the theory to a topological QFT in one higher dimension. 

This was the philosophy adopted in~\cite{Haehl:2013hoa}. As reviewed in~\cref{sec:intro} the anomaly induced parity-odd contributions to hydrodynamics are captured by an effective action which is constructed as a transgression form between a pair of connections in five dimensions. One is the bare background connection $A^\text{b}_\mu$, and  the other is the shadow hydrodynamic connection $\hat{A}_\mu = A^\text{b}_\mu + \mu\, u_\mu$, where the background gauge potential is shifted by the hydrodynamic velocity field (suitably normalized by the chemical potential). Both these connections and the hydrodynamic variables have been uplifted from four to five dimensions. The uplift of the latter is trivial, we simply pushforward the vector field $u^\mu$ from four to five dimensions. 

The transgression form when varied with respect to the background sources, reproduces correctly the anomaly induced constitutive relations. We recall that the anomalous terms in (the $4$-dimensional) hydrodynamic currents are proportional the vorticity vector $w^\mu = \frac{1}{2} \, \epsilon^{\mu \lambda \alpha \beta} u_\lambda\, \nabla_\alpha u_\beta$, with $T^{\mu\nu}_\text{anom} = -8\, c_A \, \mu^3\, w^{(\mu}\, u^{\nu)}$ and  $J^\mu_\text{anom} = -6\, c_A \, \mu^2\, w^\mu$, when all the sources are turned off.

However, this action did not respect the anomalous Ward identities, for one obtained an additional contribution proportional to the shadow connection. This was argued to be cured by passing onto a Schwinger-Keldysh construction, and the inclusion of an explicit influence functional therein.  In particular, doubling the variables to have L and R indices,  a transgression form interpolating between $\hat{A}_L$ and $\hat{A}_R$ was argued to leave the anomalous constitutive relations unaffected, but cure the Ward identity.

As we are interested in constructing an effective action parameterized by the physical hydrodynamic modes in flat spacetime, we may set $A^\text{(b)} =0$, With this restriction, the influence functional of~\cite{Haehl:2013hoa} is a transgression form between $\mu_R\, u_R$ and $\mu_L\, u_L$. A cursory examination of its vertices reveals consistency with the structure we find here (at cubic order).  This can be seen, for instance, by working with the velocity one-form, with a fixed component in the direction of inflow, and using $d u = 2\, w- u\wedge a$, with $w$ being the vorticity two-form and acceleration $1$-form, respectively. The spatial part of $u$ along the QFT directions is, roughly speaking, parameterized by $\POp$, temporal direction by $\EOp$, and the chemical potential by $\QOp$. 

However, a detailed matching is complicated by the fact that one has to perform a field redefinition to pass from our fields $\{\POp_\eta,\QOp,\EOp\}$ to the hydrodynamic chemical potential and velocity. The fact that this is necessary can be seen from the fact that the analysis of~\cite{Haehl:2013hoa} fixes a particular fluid frame, the entropy frame, since there is no anomalous contribution to the entropy current. The holographic analysis, where we decoupled the kinetic terms for the physical modes, is more natural in a different frame (akin to the Landau frame). While one could attempt to work out the detailed field redefinition, it would be instructive to find a more direct argument to tie back our results to those of~\cite{Haehl:2013hoa}.

\section*{Acknowledgements}

It is a pleasure to thank R.~Loganayagam for insightful discussions.  We would also like to thank Felix Haehl for comments on a draft. MR was supported by U.S.\ Department of Energy grant DE-SC0009999 and funds from the University of California. JV was supported by  U.S. Department of Energy grant DE-SC0020360 under the HEP-QIS QuantISED program. SZ was supported by the UC Education Abroad Program during her visit to UC Davis. 

\appendix

\section{Dynamics of the hydrodynamic modes}
\label{sec:vectors}

Letting $\vb{k} =k\, \hat{e}_x$ we can characterize the graviton and photon perturbations as follows:
\begin{itemize}[wide,left=0pt]
\item  Transverse traceless gravitons (spin-$2$ representation) can be mapped to 2 minimally coupled scalar fields~\cite{He:2021jna}. Since they  correspond to non-hydrodynamic modes, we do not consider them in our discussion.
\item Transverse vector polarization (spin-$1$) involve 2 metric and 2 gauge field modes which mix with each other and are sensitive to the anomaly. Working with chiral vector modes, we decouple these 4 fields into a pair of chiral hydrodynamic shear modes $\MX^\pm$, and a pair of non-hydrodynamic charge modes $\MY^\pm$. The dynamics is obtained along the lines described in~\cite{He:2021jna} with some changes to account for the anomaly, as outlined below. 
\item Finally, longitudinal scalars arising from the metric and the gauge field give rise to a coupled system, which can be diagonalized in terms of two fields corresponding to the energy transport $\Zd$ and charge diffusion $\Vd$ modes, respectively~\cite{He:2022deg}. 
\end{itemize} 

Retaining just the four fields corresponding to the hydrodynamic degrees of freedom, the gauge field perturbations can be parameterized as 
\begin{equation}\label{eq:Amaxpar}
\vb{A}_A^{(1)}\, dx^A 
= 
	 \frac{1}{r}\left(dv\, \Dz_+ - dr\, \dv{r}\right) \MV(r,\sk)\,\ScS +\; 
	 \vMax_\eta(r,\sk) \, \VV^\eta_i\, dx^i  \,.
\end{equation}
Here $\Dz_+$ is the derivative adapted to ingoing coordinates 
\begin{equation}\label{eq:Dzdef}
\Dz_+ = r^2f\, \partial_r + \partial_v \,,
\end{equation}	
which has a useful conjugation property to relate ingoing and outgoing modes. 
The metric perturbations are captured by the ansatz
\begin{equation}\label{eq:gABpar}
\begin{split}
ds_{(1)}^2{}
 &=  ds_{\text{vector}}^2 + ds_{\text{scalar}}^2 \,, \\
ds^2_\text{vector}
&=
		2\, r^2  \int_k\,(\vGR_{r,\eta} (r,\sk) \, dr +\vGR_{v,\eta} (r,\sk) \,dv)\,
  	\VV^\eta_i dx^i \\
ds^2_\text{scalar}
&= 	
	 \frac{\PHW}{r^{2}} \,ds_{(0)}^2 +2\,  rf\,\dv{\MW}{r} \,\frac{dr}{r^2f} \left(\frac{dr}{r^2f}-dv\right) 
     +\frac{\Dz_+\MW}{r}\, dv^2  -3f\,\PHW \left(\frac{dr}{r^2f} \right)^2.
\end{split}
\end{equation}	
Here $\ScS = e^{-i\omega\,v+i\vb{k}.\cdot \vb{x}}$ is a planar harmonic on $\mathbb{R}^{3,1}$. The vector harmonics $\VV_i^\pm$ are chiral combinations of planar harmonics with definite parity under $\vb{x} \to -\vb{x}$ and $\vb{k} \to -\vb{k}$. We give our conventions for these in~\cref{sec:harmonics}.

\paragraph{Scalar perturbations:} The functions appearing in the scalar metric perturbations are functions of $\Vd,\Zd$
\begin{equation}\label{eq:EOWMZ}
\begin{split}
\MW 
&= 
	\frac{r}{\Lk}\left[\Dz_+  -\frac{r^2\, f'}{2}\right]  \left(\frac{\Lk}{\bRQ\, h} \, \Vd + h\, \Zd \right) + \frac{6\, r^3f\, a'}{\Lk}\, \MV \,, \\
\PHW 
&= 
	   \frac{1}{ \Lk}\left[r\,\Dz_+ + \frac{k^2}{d-1} \right] \left(\frac{\Lk}{\bRQ\, h} \, \Vd + h\, \Zd \right)+ \frac{6\, r^3f\, a'}{\Lk}\, 
	   \MV\,.
\end{split}
\end{equation}	
Likewise, the field $\MV$ is itself a linear combination of the fields $\Vd$ and $\Zd$
\begin{equation}\label{eq:MZVdiagonalA}
\begin{split}
\MV 
&=  
	\left(\frac{a}{\bRQ} - r_+\, \frac{\BQT^2+2}{4}\right) \frac{\Vd}{h} +\left(  a + \frac{r_+\, \bRQ}{4} \, \BQT^2\right) \frac{h}{\Lk} \, \Zd \,.
\end{split}
\end{equation}
This parameterization and the simplification of the Einstein-Maxwell dynamics to the classical action given in~\eqref{eq:XVZbulk} was explained in detail in~\cite{He:2022deg}. This continues to be the case with the Chern-Simons term, for the reader can easily check that there is no contribution solely from the scalar sector. As explained in~\cite{Loganayagam:2022teq} the field $\MV$ is regular at the zeros of the function $\Lk$, which occur at complex momenta. This is a somewhat non-trivial statement; the fields $\MZ$ has an apparent singularity at these locations, while $\MV$ has a pole. Nevertheless, the particular combinations appearing in the gauge potential and metric functions ends up being manifestly regular. 

The normalization factors are functions of the modified spatial momenta $\BQT$ and are 
\begin{equation}\label{eq:NVZ}
\begin{split}
\Np{\Vd}(\BQT)
&= 	
	\frac{r_+^2}{8}\, \left(1+\BQT^2\right)\left(2+\BQT^2\right) , \\ 
\Np{\Zd}(\BQT)
&= 	
	\frac{3\,r_+^2}{4}\, \BQT^2 \left(1+\BQT^2\right)  .
\end{split}
\end{equation}
The potentials $\VVd$ and $\VZd$ are also obtained in the reference mentioned. While they do not play any role in our analysis, we nevertheless quote them here for completeness:
\begin{equation}\label{eq:VZdPot1}
\begin{split}
\VVd 
&= 
     \frac{2\, k^2\,(1-h)\, f}{(1+\BQT^2)\,\Lk^2}\; \VVd^{(1)} 
    +  \frac{r^3 f' f}{2\,(1+\BQT^2)\,h^2\,\Lk^2} \, \BQT^2\; \VVd^{(2)}\,,\\
\VZd 
&= 
    \frac{2\, k^2\, (1-h)\, f }{(1+\BQT^2)\,\Lk^2} \;  \VZd^{(1)} - \frac{3\, r^5f f'}{  (1+\BQT^2)\, \Lk^2\, h^2}\, \BQT^2 \; \VZd^{(2)}  \,,
\end{split}
\end{equation}
where
\begin{equation}\label{eq:VZdPot2}
\begin{split}
\VVd^{(1)} 
&=
    -\frac{4}{3}\, \Lk^2+ 6\,r^5 f' f \left(\frac{1}{h}-2\right)^2 - \BQT^2 \left(2f(h-2)+r f'\,h\right)\frac{r^2\,\Lk}{h^2}\\
&\qquad
    +2\left[\left(2-\left(1+2 \,h\right)h\right)f+(2\,h-1)\,r f'\, h\right]\frac{r^2\,\Lk}{h^2}\, ,\\
\VVd^{(2)} 
&= 
    2\left(2h-1\right)h \,\Lk^2  + 6\left[4\,f-11\,f h+\left(8\,f-r f'\right)h^2\right]
    r^2 \, \Lk\\
&\qquad
-18\,(2h-1)^2\, r^5 f' f\,,\\
\VZd^{(1)} 
&= 
    \frac{4}{3}\, \Lk^2 + 2\left[2\left(  f-r f'\right)h- (1+\BQT^2) f\right]
    \frac{r^2\,\Lk}{h} \\
&\qquad
    + \frac{6\,r^5 f f'}{h^2}\left(1+\BQT^2-2\,h\right)\left(2\,h-1\right) \,,\\
\VZd^{(2)} 
&= 
    \left(  f-\frac{1}{2}r f'\right)h\,\Lk - 3\,r^3 f f'\,(2h-1)\,.
\end{split}
\end{equation}

\paragraph{Vector sector:} We need to simplify the dynamics of the vector. We proceed this following~\cite{He:2022deg} cleverly parameterizing $\{\vGR_{v,\eta},\vGR_{r,\eta}, \vMax_\eta\}$ in terms of two fields $X_\eta$ and $Y_\eta$ as follows 
\begin{equation}\label{eq:XYpar1}
\begin{aligned}
&
\dv{\vGR_v^\eta}{r} + i\omega  \, \vGR_r^\eta  
=
	\frac{k^2}{r^5}  (X_\eta +2\,Y_\eta)  \,, && \quad 
\dv{\vGR_x^\eta}{r}+ik \,\vGR_r^\eta 
 = -
 	\frac{ik}{r^3} \, \dv{X_\eta}{r} \,,\\ 
&k\, \vGR_v^\eta -\omega \, \vGR_x^\eta 
= 
	\frac{k}{r^3}\, \Dz_+ X_\eta \,, && \quad 
\vMax_\eta 
=
	 -\frac{k^2}{2\, \mu\,r_+^{2} } \, Y_\eta \,.\\	
\end{aligned}
\end{equation}	
The equations of motion arising from~\eqref{eq:SEMax} can be shown to be equivalent to the following two coupled equations:
\begin{equation}\label{eq:XYcoupled}
\begin{split}
&
  \mathfrak{D}_{-3} X_{\eta} 
=
    2\, k^2 f \,  Y_{\eta} \,,\\
&
    \mathfrak{D}_1 Y_{\eta }
= 
    4\, a^2\,f\, (X_{\eta} + 2 \,Y_{\eta} ) - 8\, \eta\, \kcs\, r\, f\, a' \, Y_{\eta }\,.
\end{split}
\end{equation}  
The operator $\mathfrak{D}_\ann$ is defined to be 
\begin{equation}\label{eq:Dopdef}
\mathfrak{D}_\ann \equiv \frac{1}{r^\ann}\, \Dz_+ (r^\ann\, \Dz_+) + (\omega^2 - k^2\, f)\,.
\end{equation}	

These equations in turn can be decoupled by the  functional linear combination
\begin{equation}\label{eq:CXYdef}
\begin{split}
X_\eta 
&=
	 -\left(\BQTCS^2+2 \left[1+\frac{4\,\eta\, \kcs\, \bqt}{\bRQ}\right] \right) \, \MX_\eta + \BQTCS^2\, \frac{h}{1-h} \, \MY_\eta\,,\\
Y_\eta 
&=
	 (1-h) \,\MX_\eta + h\, \MY_\eta \,,
\end{split}
\end{equation}
The deformed momentum parameter $\BQTCS$ which appears courtesy the basis rotation coefficients when we decouple the equations and is given in~\eqref{eq:BQTdefs}. The decoupled equations of motion themselves are 
\begin{equation}\label{eq:XYdecouple}
\begin{split}
\frac{1}{r\,(1-h)^2} \, \Dz_+\left(r\, (1-h)^2\, \Dz_+ \MX_\eta\right)
 + r_+^2 \left(\bwt^2 - \bqt^2 \,f+ \bRQ^2\, (1-h)\, \BQTCS^2\, f\right) \MX_\eta  
&
	=0 \,,\\ 
\frac{1}{r\,h^2} \, \Dz_+\left(r\, h^2\, \Dz_+ \MY_\eta\right)
 + r_+^2 \left(\bwt^2 - \bqt^2 \,f+ \bRQ^2\, (1-h)\, \left(\BQTCS^2 + 8\,\eta\, \kcs\, \frac{\bqt}{\bRQ}\right) f\right) \MY_\eta  
&
	=0\,. \\ 	
\end{split}
\end{equation}	
It can easily be checked that setting $\kcs =0$ one recovers the equations for the Einstein-Maxwell system analyzed in~\cite{He:2021jna}.

One can similarly plug in the parameterization of the metric and gauge field in terms of $\MX$ and $\MY$ into~\eqref{eq:SEMax} and obtain the off-shell action parameterized by these variables. The procedure parallels the Einstein-Maxwell discussion reported in~\cite{He:2021jna}. One finds 
after some algebra the final action 
\begin{equation}
S_{\text{vector}} = S_{\MX} +S_{\MY} + S_{\MX\MY, \text{bdy}} + \frac{1}{2}\int_k \frac{1}{r^5}\left(\frac{\mathcal{E}_X}{f}\right)^2\,.
\end{equation}
The bulk part of the action is completely diagonalized in the variables $\MX, \MY$ and takes the form
\begin{equation}\label{eq:Sdiag}
\begin{split}
S_{\MX} 
&=
	-\int_k \int dr  \, k^2\,  \frac{\sqrt{-g}}{r^6} \,\Np{\MX,\eta}
	\left[
		\nabla^{A}\MX_\eta\,\nabla_{A}\MX_\eta
		-\BQTCS^2\, \bRQ^2\, \frac{r_+^2}{r^2}\, (1-h)\, \MX_\eta\,\MX_\eta
		\right]\,,\\
S_{\MY} 
&=
	-\int_k \int dr \, k^2\,   \frac{h^2\,\sqrt{-g}}{r^2} 
	 \,\Np{\MY,\eta} \left[\nabla^{A}\MY_\eta\, \nabla_{A}\MY_\eta 
		+\left(\BQTCS^2 +8\, \eta\, \kcs\, \frac{\bqt}{\bRQ}\right)
		\bRQ^2\, \frac{r_+^2}{r^2}\, (1-h)\MY_\eta\MY_\eta\right]\,,
\end{split}
\end{equation}
The normalization factors are functions of momenta depending on the chirality
\begin{equation}\label{eq:NormCon}
\begin{split}
\Np{\MX,\eta}
&= 	
	\left(\BQTCS^2+1 +4\,\eta\,\kcs\, \frac{\bqt}{\bRQ}\right)\left(\BQTCS^2+2 +8\, \eta \, \kcs\, \frac{\bqt}{\bRQ}\right)\,,\\ 
\Np{\MY,\eta}
&= 
	\frac{1}{\RQ^4} \, \BQTCS^2
		\left(\BQTCS^2+1 +4\,\eta\,\kcs \,\frac{\bqt}{\bRQ}\right)\,.
\end{split}
\end{equation}

The boundary term itself is not quite diagonal, and is given by 
\begin{equation}
\begin{split}
L_{\MX\MY,\text{bdy}} 
&= 
	-\left(\BQTCS^2+ 2 +8\,\eta\,\kcs\, \frac{\bqt}{\bRQ}\right)^2 
	k^2\, \pi_{\MX,\eta} \, \MX_\eta 
 	- \left(\frac{\BQTCS^2}{\RQ^2}\right)^2 k^2\, 
 	\pi_{\MY,\eta} \, \MY_\eta
 	+\frac{2\,k^2\,\bqt^2}{\bRQ^2} \, \frac{f(r)}{r^2}\MX_\eta\, \MX_\eta\\ 
&\qquad 
 	+\frac{k^4}{\BQTCS^2}\, f h
 	\left(\frac{h}{1-h}\,\BQTCS^2-2-8\,\eta\,\kcs\, 
 	\frac{\bqt}{\bRQ}\right)\MY_\eta \MY_\eta 
 	-\frac{4\,k^2\, \bqt^2}{\bRQ^2}\,\frac{f}{r^2}\, \MX_\eta\, \MY_\eta\\ 
 &\qquad 
	 + \frac{k^2}{\RQ^4}\, \BQTCS^2\left(\BQTCS^2+2 +8\,\eta\,\kcs\, \frac{\bqt}{\bRQ}\right)
	 \left(\frac{1-h}{h}\, \MX_\eta\, \pi_{\MY,\eta} + \frac{h}{1-h}\,\RQ^4\,\MY_\eta\, \pi_{\MX,\eta} \right)\,.
\end{split}
\end{equation}
We introduced the conjugate momenta
\begin{equation}\label{eq:ConjMDiag}
\pi_{\MX,\eta} = -r^{-3}\Dz_+\MX_{\eta}\,, \qquad 
\pi_{\MY,\eta} = -r\, h^2\, \Dz_+ \MY_{\eta}\,.
\end{equation}
The boundary conditions on the fields $\MX_\eta$ and $\MY_\eta$ are Neumann and Dirichlet, respectively (for purposes of computing correlation functions). This is consistent both with the boundary conditions inferred from the above boundary action (evaluated asymptotically), and with the fact that the fields have an auxiliary dilaton dressing in their kinetic term. The field $\MX_\eta$ has a dilaton $r^{-6}$ which is a non-Markovian field of index $\ann=-3$ as in~\cite{He:2021jna}. On the other hand, $\MY_\eta$ is Markovian as its dilatonic modulation is $r^{-2}\,h^2$.

\paragraph{Cubic vertex:} The cubic vertex arising from the Einstein-Maxwell-Chern-Simons action~\eqref{eq:SEMax} is quite complicated owing to the contributions from the Einstein-Maxwell sector. This parity-even part involves a cubic vertex with three scalar modes, viz., $\Zd^a\, \Vd^{3-a}$ for $a\in \{0,1,2,3\}$ from the gravitational part. But it also has interactions between mixing the vector and scalar sectors of the perturbations: interactions both the form $\MX_\eta\, \MX_\eta\, \Zd$,  $\MX_\eta\, \MX_\eta\, \Vd$, and $\MX_\eta\, \Zd^{2-b}\, \Vd^b$ for $b \in \{0,1,2\}$, are in principle admissible.  We will not evaluate the explicit form of these vertices here. 

Rather, we will focus on the part of the cubic vertex arising from the Chern-Simons term. This part only receives contribution from the gauge field~\cref{eq:Amaxpar}. However, since the potential receives contribution from both the vector and scalar sectors, we will end up with terms mixing the two sectors. The main difference is in the polarization pattern, and the spatial momentum dependence, which are dictated by the form of the Chern-Simons contraction. We have found it convenient to adopt a basis of form factors labeled by $4$-momenta and polarization, introduced in~\cref{sec:harmonics}, cf.,~\eqref{eq:VVVffactor}-\eqref{eq:VSSffactor}. The result quoted in the main text was obtained by directly evaluating the Chern-Simons term with the parameterization~\cref{eq:Amaxpar} (setting $\MY =0$).

\section{Solutions to the equations of motion}\label{sec:vecXYsols}

We present here the solutions to the  equations~\cref{eq:XYdecouple} in a gradient expansion for small frequencies and momenta. As in~\cite{He:2021jna,He:2022deg} we parameterize the ingoing Green functions as:
\begin{equation}\label{eq:GinExp}
\begin{split}
\Gin{\MX,\eta}(r) 
&=
	 \exp\left(\sum_{n,m=1}^{\infty}(-i)^n \, \xser{n,m}(r) \, \bwt^n \, \bqt^m\right)\,, \\ 
\Gin{\MY,\eta}(r) 
&= 
	\exp\left(\sum_{n,m=1}^{\infty}(-i)^n \, \yser{n,m}(r)\, \bwt^n \, \bqt^m
		\right)\,.
\end{split}
\end{equation}
Notice that unlike the analysis of~\cite{He:2021jna} where the sums only included even powers of momentum, in the present discussion owing to the parity breaking, we must include also odd powers due to the Chern-Simons contribution to the potential. In writing the solutions, we find it useful to introduce the dimensionless ratio
\begin{equation}\label{eq:sdcdef}
\sdc = \frac{\RQ^2}{r_+^2} \,.
\end{equation}	

We should note here that we are not presenting here the solutions for the fields $\Vd$ and $\Zd$. These can be found in Appendix C of~\cite{He:2022deg}, where they are given for arbitrary dimension. Specialization to $d=4$ is straightforward, with the caveat that the leading power law divergence becomes logarithmic (which can however be understood using dimensional regularization).

\subsection{Solution for \texorpdfstring{$\MX$}{X field}}\label{sec:Xsols}

The solutions for the field $\MX$  are mildly different from the ones found in the aforementioned reference. In particular, all frequency-only solutions are of the form:
\begin{equation}\label{eq:Xsol1}
\xser{n,0}(r) = \Mser{-3}{n,0}(r)\,,
\end{equation}
where $\Mser{-3}{n,m}(r)$ are the corresponding functions for a non-Markovian scalar of index $\ann=-3$ in $d=4$. From the structure of the potential, we can see that the functions associated with a single power of momentum all vanish
\begin{equation}\label{eq:Xsol2}
\xser{n,1}(r) =0\,.
\end{equation}

The gradient expansion coefficients with at least two powers of momentum receive two kinds of modifications:  a correction proportional to $\sdc$, related to the coupling between the metric and the gauge field arising from the Maxwell action, and a term proportional to $\eta\,\kcs$, resulting from the Chern-Simons coupling. The explicit expressions are:
\begin{equation}\label{eq:Xsol3}
\begin{split}
\xser{0,2}(r) 
&= 
	\Mser{-3}{0,2}(r) + \frac{\sdc}{6}\,\Dfn{3}{2,0}(r)\,,\\
\xser{1,2}(r) 
&= 
	\Mser{-3}{1,2}(r) - \frac{\sdc}{3}\,\Mser{-3}{2,0}(r)\,,\\
\xser{0,3}(r) 
&= 
	\frac{-2\,\sdc \,\eta\,\kcs}{3\,\bRQ} \, \Dfn{3}{2,0}(r)\,,\\ 
\xser{2,2}(r) 
&= 
	\Mser{-3}{2,2}(r) - \frac{\sdc}{3}\left(\Mser{-3}{3,0}(r)+\frac{1}{2}\,\Dfn{-3}{3,0}(r) \right)\,,\\
\xser{1,3}(r) 
&= 
	\frac{4\,\sdc \,\eta\,\kcs}{3\,\bRQ}\,\Mser{-3}{2,0}(r)\,,\\ 
\xser{0,4}(r) 
&= 
	\Mser{-3}{0,4}(r) - \frac{\sdc}{12\,\bRQ^2}\Dfn{3}{2,0}(r) - \frac{1}{3\,\bRQ^2}\Dfn{\MX}{a}(r)+\sdc\Dfn{\MX}{b}(r) + \frac{8\sdc\,\kcs^2}{3\, \bRQ^2}\Dfn{3}{2,0}(r \,.
\end{split}
\end{equation}
We have introduced a few special functions, which themselves are 
\begin{equation}\label{eq:DeltaX}
\begin{split}
\Dfn{3}{2,0}(r) 
&= 
	\Fint{r^{-3}-r^3} ,\\ 
\Dfn{-3}{3,0}(r) 
&= 
	\Fint{r^{3} \left(\Dfnh{3}{2,0}\right)^2}\,,\\ 
\Dfn{\MX}{a}(r) 
&= 
	\int^r du\, u\,\Dfn{3}{2,0}(u)\,,\\
\Dfn{\MX}{b}(r) 
&= 
	\int^r du\, \frac{u}{f(u)} \int^{u}_{1}du'\left[\frac{f'(u')\,\Dfn{3}{2,0}(u')}{12\,(1+Q^2)}
	+\frac{\sdc}{36} \, \frac{f(u')}{u'} \left(\frac{d}{du'}\Dfn{3}{2,0}(u')\right)^2\right].
\end{split}
\end{equation}
We have written the expressions introducing a shorthand $\Fint{g}$ for the integral transform
\begin{equation}\label{eq:Fint}
\Fint{g}(r) = \int_{r_+}^{r}\, \frac{dr'}{r'^2\,f(r')}\, g(r')\,,
\end{equation}
For functions, we use $\hat{g}(r)$ to denote the situation where the zero point is set at the horizon, viz.,  $\hat{g}(r_+)=0$.

\subsection{Solution for \texorpdfstring{$\MY$}{Y field}}\label{sec:Ysols}

Unlike $\MX$, solutions for $\MY$  are heavily modified by the Chern-Simons term. The corrections appear at all powers of momentum starting at the linear order. We present the answers using the integral transform
\begin{equation}\label{eq:Hint}
\FHint{g}(r) = \int_{r_+}^{r} \frac{dr'}{r'^2\,h(r')^2\,f(r')}\, g(r').
\end{equation}

The simplest solutions are those for zero momentum. These frequency expansion coefficients, do not get any  correction from the potential of the bulk wave equation. Explicitly, one has 
\begin{equation}\label{eq:Ysol1}
\begin{split}
\yser{1,0}(r) 
&= 
	-\Fint{1-\frac{h(r_+)^2}{h(r)^2\, r}}\,,\\ 
\yser{2,0}(r) 
&= 
	-h(1)^2\, \FHint{r^{-1}\, \Dfnh{\MY}{2,0}(r)}\,,\\ 
\yser{3,0}(r) 
&= 
	-2\,h(1)^2 \, \FHint{r^{-1}\, \yserh{2,0}(r)}\,,\\ 
\yser{4,0}(r) 
&= 
	-2\,h(1)^2\, \FHint{r^{-1} \left(\yserh{3,0}(r)+\frac{1}{2}\Dfnh{\MY}{3,0}(r)\right)}\,.
\end{split}
\end{equation}
Here we introduced
\begin{equation}\label{eq:DeltaY1}
\begin{split}
\Dfn{\MY}{2,0}(r) 
&= 
	\Fint{\frac{h(r_+)^2}{h(r)^2\,r}-\frac{h(r)^2}{h(r_+)^2}\, r} \,,\\
\Dfn{\MY}{3,0}(r) 
&= 
	h(r_+)^2\, \FHint{r^{-1}\, \Dfnh{\MY}{2,0}(r)^2 }\,.
\end{split}
\end{equation}
The functions are identical to the ones introduced in~\cite{He:2021jna}, up to a change of coordinates.

The solutions correcting the momentum dependence are more elaborate, but can be simplified using some identities specific to $d=4$. The functions of interest are 
\begin{equation}\label{eq:solY2}
\begin{split}
\yser{0,1}(r) 
&= 
	\frac{16\,\eta\,\kcs \bRQ}{12} \, 
	\FHint{r^{-1}\left(h(r)^3-h(r_+)^3\right)}\,,\\ 
\yser{1,1}(r) 
&= 
	-2h(r_+)^2\, \FHint{r^{-2}\yserh{0,1}(r)} \,,\\ 
\yser{0,2}(r) 
&= 
	\FHint{\frac{\log r}{r}} - \frac{\sdc}{24} \,\Dfn{\MY}{0,2}(r) + \frac{16}{9}\,\kcs^2\bRQ^2 \,\FHint{\frac{1}{r}\,\Dfnh{\kappa}{0,2}(r)}\,,\\
\yser{2,1}(r) 
&= 
	-h(r_+)^2\, \FHint{r^{-1}\left(\yserh{1,1}(r)-2\yserh{0,1}(r)\Dfn{\MY}{2,0}(r)\right)}
	+2\,h(r_+)^2\, \FHint{r^{-1}\Dfnh{\MY}{2,1}(r)}\,,\\ 
\yser{1,2}(r) 
&= 
	-2\,h(r_+)^2\, \FHint{r^{-1}\left(\yserh{0,2}(r)-\yserh{0,1}(r)^2\right)}\,,\\
\yser{0,3}(r) 
&= 
	16\,eta\,\kcs\sdc\bRQ\,\FHint{r^{-1}\Dfnh{\MY}{0,3}(r)} - \frac{1}{2\,\bRQ^2}\yser{0,1}(r) - 2 \int_{0}^r du \, \yser{0,2}(u)\frac{d}{du}\yser{0,1}(u)\,,\\ 
\yser{3,1}(r) 
&= 
	-2\,h(r_+)^2\, \left(\FHint{r^{-1}\yserh{2,1}(r)}
	+2 \,h(r_+)^2\,\FHint{r^{-1}\Dfnh{\MY}{3,1}(r)}\right),\\ 
\yser{2,2}(r) 
&= 
	-2\,h(r_+)^2\, \FHint{r^{-1}\left(\yserh{1,2}(r)+\Dfnh{\MY}{2,2}(r)\right)}\,, \\
\yser{1,3}(r) 
&= 
	-h(r_+)^2\, \FHint{r^{-1} \left(2\yserh{0,3}(r)+\Dfnh{\MY}{1,3}(r)\right)}\,, \\ 
\yser{0,4}(r) 
&= 
	\FHint{r^{-1}\Dfnh{\MY}{0,4}(r)}+\frac{2\eta\,\kcs}{\bRQ^{3/2}}\,\yser{0,1}(r)\,.
\end{split}
\end{equation}
The auxiliary functions introduced to write these solutions compactly are themselves given as 
\begin{equation}\label{eq:DeltaY2}
\begin{split}
\Dfn{\MY}{0,2}(r) 
&= 
	\FHint{r^{-1}\left(4\left(\frac{h(r)^2}{r^2}-h(r_+)^2\right)-2\left(\frac{h(r)^2}{r^2}-h(r_+)\right)+14 \left(1-\frac{1}{r^2}\right)\right)}\,,\\ 
\Dfn{\kappa}{0,2}(r) 
&= 
	\FHint{r^{-1}\left(h(r)^3-h(1)^3\right)^2}\,,\\ 
\Dfn{\MY}{2,1}(r) 
&= 
	\frac{1}{h(r_+)^2}\, \Fint{h(r)^2\,r^{d-3}\,\yserh{0,1}(r)}\,,\\ 
\Dfn{\MY}{0,3}(r) 
&= 
	\int_{r_+}^{r}\, du \, h(u)^2 u^{-3}\, \yser{0,2}(r)\,,\\ 
\Dfn{\MY}{3,1}(r) 
&= 
	\FHint{r^{-1}\,\yserh{0,1}(r)\,\Dfnh{\MY}{2,0}(r)} + 
	\frac{4\eta\,\kcs\bRQ}{3\,h(r_+)^2}\, 
	\FHint{r^{-1}(h(r_+)^3-h(r)^3)\,\yserh{2,0}(r)}\,,\\ 
\Dfn{\MY}{2,2}(r) 
&= 
	2\,h(r_+)^2\, \FHint{r^{-1}\yserh{0,1}(r)^2} - \int_{r_+}^{r}\,du\,
		\Dfnh{\MY}{2,0}(r)\,\frac{d}{du}\yser{0,2}(u)\\
&\qquad
	-\int_0^{r}du\,\left(\yserh{1,1}(r)-2\Dfnh{\MY}{2,1}(u)
	-2\,\yserh{0,1}(u)\Dfnh{\MY}{2,0}(u)\right)\frac{d}{du}\yser{0,1}(u)\,, \\ 
\Dfn{\MY}{0,4}(r) 
&= 
	-\int_{0}^{r} du\, h(u)^2\, f(u)\, u^{3}\left(\frac{d}{du}\yser{0,2}(u)\right)^2 + \frac{\sdc}{12\,\bRQ^2}\frac{1+h(r)+h(r)^2}{r^2}\\ 
&\qquad 
	+\frac{8}{3}\eta\,\kcs\bRQ \, \int_{r_+}^{r}\, du\,\left(h(r_+)^3-h(r)^3\right)\frac{d}{du}\yser{0,3}(u)\,.
\end{split}
\end{equation}
Setting $\kcs =0$ in these expressions recovers the parity preserving solutions of the Einstein-Maxwell theory~\cite{He:2021jna}.  

\subsection{Asymptotics and dispersion relation}\label{sec:asdisp}

The asymptotic behavior of the functions $\Gin{\MX}(r,\bwt,\bqt)$ and $\Gin{\MY}(r,\bwt,\bqt)$ in the limit $r\rightarrow \infty$ can be determined given the explicit solutions of the two preceding subsections.

For the non-Markovian field  $\MX$, which has the minimal modifications due to the Chern-Simons coupling, we find
\begin{equation}\label{eq:XAsymp}
\begin{split}
\Gin{\MX,\eta} 
&= 
	1 + \Kin{\MX}(\bwt,\bqt)\, \frac{r^2}{2} +\cdots\,, \\
\Kin{\MX,\eta}(\bwt,\bqt) 
&= 
	\Kin{\MX}(\bwt,\bqt)\big|_{\kcs=0} + \frac{2\sdc}{3\, \bRQ} 
	\eta\,\kcs\, \bqt^3\left(1+2\,i\,\Dfn{3}{2,0}(r_+)\,\bwt
		-\frac{4\eta\,\kcs}{\bRQ}\bqt\right)\,,
\end{split}
\end{equation}
where the ellipsis stands subleading divergences as $r\rightarrow\infty$ that are canceled by the standard counterterms, as well as terms that have a finite limit.  The  piece $\Kin{\MX}(\bwt,\bqt)\big|_{\kcs=0}$ is the same function found in~\cite{He:2021jna} in the absence of a Chern-Simons coupling. The explicit expression 

On the other hand, for the Markovian field $\MY$ we find 
\begin{equation}\label{eq:YAsymp}
\begin{split}
-\pi_{\MY,\eta} 
&= 
	r \, h(r)^2\, \Dz_+\Gin{\MY} 
	= \Kin{\MY}(\bwt,\bqt) +\cdots\,,\\ 
\Kin{\MY,\eta}
&=
	-i\, h(r_+)^2\,\bwt + \frac{4}{3}\,\sdc\bRQ\,\eta\,\kcs
	\left(3\,h(r_+)+\sdc^2\right)\,\bqt -h(r_+)^2\, \Dfn{\MY}{2,0}(r_+)\bwt^2 \\
&\quad
	-2\,i \,h(r_+)^2\,  \yser{0,1}(r_+)\,\bwt\,\bqt 
	+ \left(\frac{\sdc}{2}\left(-1-\frac{\sdc}{2}+\frac{\sdc^2}{3}\right)
	+\frac{16}{9}\,\kcs^2\,\Dfn{\kappa}{0,2}(r_+)\right) \bqt^2 \\
&\quad
	+ 2\,i\, h(r_+)^2\yser{2,0}(r_+)\,\bwt^3 
	-h(r_+)^2 \left(\yser{1,1}(r_+)+2\,\yser{0,1}(r_+)\, \Dfn{\MY}{2,0}(r_+)
	-2\Dfn{\MY}{2,1}(r_+)\right)\bwt^2\bqt \\ 
&\quad
 	+ 2\, i\,  h(r_+)^2 \left(\yser{0,1}(r_+)^2 + \yser{0,2}(r_+)\right)
 	\bwt\bqt^2 
 	- \frac{2\sdc}{3\,\bRQ}\,\eta\,\kcs\left(3\, h(r_+)^2+\sdc^2+24\,\bRQ^2\,\Dfn{\MY}{0,3}(r_+)\right)\bqt^3\\ 
&\quad 
	+ h(r_+)^2 \left(2\, \yser{3,0}(r_+)+\Dfn{\MY}{3,0}(r_+)\right) \bwt^4
	+2\,i\, h(r_+)^2 \left(\yser{2,1}(r_+)
	+ 2\, h(r_+)^2\, \Dfn{\MY}{3,1}(r_+)\right)\bwt^3\bqt\\ 
&\quad
	-2\, h(r_+)^2\left(\yser{1,2}(r_+) + 
	\Dfn{\MY}{2,2}(r_+)\right)\bwt^2\bqt^2 
	- i\, h(r_+)^2\left(2\,\yser{0,3}(r_+) + \Dfn{\MY}{1,3}(r_+)\right)
		\bwt \bqt^3\\ 
&\quad 
	-\left(\Dfn{\MY}{0,4}(r_+)-\frac{8\,\sdc}{3\,\bRQ^2}\,\kcs^2\left(3\,h(r_+)^2+\sdc^2\right)\right)\bqt^4 \,.
\end{split}
\end{equation}
This Markovian Green's function receives corrections from the Chern-Simons coupling at all orders in momentum. The reader can check that this function reduces to its corresponding quantity in~\cite{He:2021jna} when we set $\kcs=0$\footnote{Notice that all functions of the form $\yser{n,m}(r_+)$ with $m$odd are proportional to $\eta\,\kcs$.}. We have again elided over the pieces that should be cancelled by counterterms. In fact, for $d=4$ the only divergent contribution is of the form $\left(-\bwt^2+\bqt^2\right)\log r$.

\subsection{Dispersion Relations}\label{sec:dispersionfns}

We compile here the analytic expressions for the diffusive dispersion relations, for chiral shear waves, and for charge diffusion. The former simplifies~\eqref{eq:XAsymp}, while the latter has been directly taken from~\cite{He:2022deg} and specialized to $d=4$. 

For the shear diffusion, the coefficients in the gradient expansion of $\mathfrak{h}_{m,n}^{s,\eta}$ appearing in~\eqref{eq:hdispcfs} are 
\begin{equation}
\begin{split}
\mathfrak{h}^{\scriptstyle{s,\eta}}_{3,0} 
&= 
	\Dfn{3}{2,0}(r_+)^2-2\, \Mser{3}{2,0}(r_+)\,, \qquad 
	\mathfrak{h}^{\scriptstyle{s,\eta}}_{1,2} = -\Mser{3}{0,2}(r_+)	
	+\frac{1}{2}\left(1-\frac{2}{3}\, \frac{\RQ^2}{r_+^2}\right) 
	\Dfn{3}{2,0}(r_+)\,, \\ 
\mathfrak{h}^{\scriptstyle{s,\eta}}_{0,3} &= \frac{2}{3}\,\frac{\eta\,\kcs}{\bRQ}\,\frac{\RQ^2}{r_{+}^{2}}\,, \qquad 
\mathfrak{h}^{\scriptstyle{s,\eta}}_{4,0} = \Dfn{3}{2,0}(r_+)^3-2\Mser{3}{3,0}(r_+)-\Dfn{3}{3,0}(r_+)-4\Mser{3}{2,0}(r_+)\Dfn{3}{2,0}(r_+)\,, \\ 
\mathfrak{h}^{\scriptstyle{s,\eta}}_{2,2} &=-\frac{1}{2}\left[2\Mser{3}{1,2}(r_+)+\Mser{3}{2,0}(r_+)+2\Dfn{3}{1,2}(r_+)+4\Mser{3}{0,2}(r_+)\Dfn{3}{2,0}(r_+)-\frac{3}{2}\Dfn{3}{2,0}(r_+)^3\right. \\ 
&\quad 
\left.
+\frac{1}{3}\frac{\RQ^2}{r_+^2}\left(3\Dfn{3}{2,0}(r_+)^2-4\Mser{3}{2,0}(r_+)\right)\right]\,, \\
\mathfrak{h}^{\scriptstyle{s,\eta}}_{1,3} &=-\frac{4}{3}\,\frac{\RQ^2}{r_{+}^{2}}\,\frac{\eta\,\kcs}{\bRQ}\, \Dfn{3}{2,0}(r_+)\,, \\ 
\mathfrak{h}^{\scriptstyle{s,\eta}}_{0,4} &=
\frac{1}{4}\frac{\RQ^2}{r_+^2}\left(\frac{1}{10}\frac{\RQ^2}{r_+^2}+\frac{1}{3}\frac{1}{\bRQ^2}-\frac{3}{8}\right)
-\frac{1}{16}\left(2\Mser{3}{0,2}(r_+)+4\,\Mser{5}{0,2}(r_+)-\Dfn{3}{2,0}(r_+)\right)\\ 
&\quad 
 - \frac{2}{3}\,\frac{\RQ^2}{r_{+}^{2}} \left(\frac{2\,\eta\,\kcs}{\bRQ}\right)^2\,.
\end{split}
\end{equation}
These functions are numerically evaluated and displayed in~\cref{fig:KxCoeffs}.
We should however note that most  of these coefficients have analytic expressions in terms of polylogs, but the charge dependence makes them complicated. In addition to $\Dfn{3}{2,0}(r_+) $ and $\Dfn{\MY}{2,0}(r_+)$, one other function which has a simple expression is 
\begin{equation}\label{eq:SpFunc}
\begin{split}
\Mser{3}{0,2}(r_+) &= -\frac{\coth^{-1}\left(\sqrt{1+4\,Q^2}\right)+\tanh^{-1}\left(\frac{2\,Q^2-1}{\sqrt{1+4\,Q^2}}\right)}{2\sqrt{1+4\,Q^2}}\,.
\end{split}
\end{equation}
The coefficients appearing in the charge diffusion relation are 
\begin{equation}\label{eq:ChargeExp}
\begin{split}
\mathfrak{h}^{\scriptstyle{c}}_{3,0} 
&= 
	\Dfn{\MY}{2,0}(r_+)^2-2\Mser{\MY}{2,0}(r_+)\,, 
	\qquad 
	\mathfrak{h}^{\scriptstyle{c}}_{1,2} =2\,\Mser{\Vd}{0,2}(r_+)\,, \\  
\mathfrak{h}^{\scriptstyle{c}}_{4,0} 
&= 
	\Dfn{\MY}{2,0}(r_+)^3-2\,\Mser{\MY}{3,0}(r_+)-\Dfn{\MY}{3,0}(r_+)
	-4\,\Mser{\MY}{2,0}(r_+)\, \Dfn{\MY}{2,0}(r_+)\,, \\ 
\mathfrak{h}^{\scriptstyle{c}}_{2,2} 
&=
	2\left(\Mser{\Vd}{1,2}(r_+)-h(r_+)^2\, \Dfn{\Vd}{2,2}(r_+)\right)\,, \\ 
\mathfrak{h}^{\scriptstyle{c}}_{0,4} 
&=
	h(r_+)^2\, \Dfn{\Vd}{0,4}(r_+)+\frac{Q^2}{3\,(1+Q^2)^2}\,\Xi_{\Vd}(r_+)\,.
\end{split}
\end{equation}
These have been evaluated numerically and plotted in~\cref{fig:KcCoeffs}.

\section{Chiral plane wave harmonics}\label{sec:harmonics}

We give a quick summary of the conventions for harmonic decomposition on $\mathbb{R}^{3,1}$, which are used in our analysis (see also Appendix F of~\cite{Ghosh:2020lel}). All of our harmonic functions $\{\ScS,\VV^\ai_i \}$ are  plane waves They are eigenfunctions of $\{i\,\pdv{v} ,-i\pdv{x^i}\}$ with eigenvalue  $\{\omega, k_i\}$.  For notational convenience we indicate both the position and momentum labels in the arguments; these are dropped in the main text for clarity.

The \emph{scalar harmonics} are simply scalar plane waves (nb: $\sk = (\omega,\vb{k})$ and $x = (v,\vb{x})$)
\begin{equation}\label{eq:sharm}
\ScS(\sk|x)\equiv e^{i \, \bk\cdot \bx-i\, \omega v}\,.
\end{equation}	
We in addition need \emph{vector harmonics}, which are transverse to the momentum vector $\vb{k}$, and satisfy $k_i\, \VV_i=0$.  There are two such vectors in three spatial dimensions. We can pick them to have definite parity under the simultaneous reversal of space and spatial momentum,  $\vb{x} \to -\vb{x}$ and $\vb{k} \to -\vb{k}$. These we refer to as the harmonic vectors $\VV^\text{E}_i$ and $\VV^\text{O}_i$, with the superscripts indicating even and odd parity. Singling out the $\hat{e}_1$ direction, they are explicitly given as~\cite{Gubser:2007nd} 
\begin{equation}\label{eq:EOvector}
\begin{split}
\VV^\text{E}(\sk|x) 
&= 
	\left( \frac{k_\perp^2}{k\,k_\perp}\, \hat{e}_1 - \frac{k_1\,k_2}{k\,k_\perp}\, \hat{e}_2 - \frac{k_1\,k_3}{k\,k_\perp}\, \hat{e}_3 \right) \ScS(\sk|x)  \,,\\
\VV^\text{O}(\sk|x) 
&=  
	\left( -\frac{k_3}{k_\perp}\, \hat{e}_2 + \frac{k_2}{k_\perp}\, \hat{e}_3 \right)  \,\ScS(\sk|x) \,.
\end{split}
\end{equation}	
Here $k$ is as usual the magnitude of the momentum, while $k_\perp^2 = k^2 - k_1^2$, given our preferred direction choice. 

We will be interested in chiral combinations of these vector harmonics, and so define
\begin{equation}\label{eq:chiralvector}
\begin{split}
\VV^\eta_i(\sk|x) 
&= 
	\frac{1}{\sqrt{2}} \left(
	\VV^\text{E}_i(\sk|x) + i\,\eta\, \VV^\text{O}_i(\sk|x)\right),\\
&=
	\frac{1}{\sqrt{2}\, k_\perp}\left( \frac{k_\perp^2}{k}\, \hat{e}_1 - \left[
	\frac{k_1\,k_2}{k} + i\, \eta\, k_3\right] \hat{e}_2 -\left[
	\frac{k_1\,k_3}{k} - i\,\eta\, k_2\right] \hat{e}_3 \right) \ScS(\sk|x)  \,.	
\end{split}
\end{equation}	
We use $\eta $ is the chirality eigenvalue, with 
\begin{equation}
\eta = 
\begin{cases}
& +1 \,, \qquad +\;  \text{chirality} \\
& -1\,, \qquad - \; \text{chirality}\,,
\end{cases}
\end{equation}	
A useful identify that follows directly from~\eqref{eq:EOvector} is
\begin{equation}\label{eq:chiralcurl}
\epsilon^{ijk} \partial_j\,\VV_k^\eta(\sk|x)
= - k\, \eta\, \delta^{il} \, \VV_l^\eta(\sk|x)\,.
\end{equation}  

These harmonics are orthonormal with respect to the flat measure, defined as
\begin{equation}\label{eq:L2norm}
\expval{\mathcal{P}(\sk|x) ,\mathcal{Q}(\sk'|x) } = 
\int\, d^4 x\, \mathcal{P}(\sk|x) \,\mathcal{Q}(\sk'|x) \,.
\end{equation}	
For the scalar and chiral vector harmonics we have therefore,
\begin{equation}
\begin{split}
\expval{ \ScS(\sk|x) ,\ScS(\sk'|x) }
 	 &=  {(2\pi)}^4\, \delta^4(\sk+\sk') \,, \\
\sum_{i=1}^3 \expval{ \VV^{\eta_1}_i(\sk|x) , \VV^{\eta_2}_i(\sk'|x) }
 &  
 	= {(2\pi)}^4\,\delta_{\eta_1\eta_2}\,  \delta^4(\sk+\sk')\,.
 \end{split}
\end{equation}
We have adopted a standard shorthand
\begin{equation}\label{eq:dxshort}
d^4 x \equiv dv \, d^3\bx \,, \qquad 
\delta^4(\sk+\sk') =  \delta(\omega+\omega')\times  \delta^{3}(\bk+\bk')\,.
\end{equation}	

In the Chern-Simons vertex we find contributions from a triplet of fields. These can arise from either the scalar or from the vector modes of the gauge potential. We will organize these as tensors in the polarization space (essentially counting the number of vectors). 

First, we have a form factor which arises from three chiral vector fields, given as
\begin{equation}\label{eq:VVVffactor}
\begin{split}
\VVV^{\eta_1\eta_2\eta_3}(\sk_1,\sk_2,\sk_3)
& = 
	\int\, d^4x\, \epsilon^{ijk}\, \VV^{\eta_1}_i(\sk_1|x)\,  \VV^{\eta_2}_j(\sk_2|x)\, \VV^{\eta_3}_k(\sk_3|x) 
\end{split}
\end{equation}	
Next, we have a contribution involving two chiral vectors and one scalar. While there are naively two independent contractions, they turn out to be related, upon using momentum conservation. These are
\begin{equation}\label{eq:VVSffactor}
\begin{split}
\VVSa^{\eta_1\eta_2}(\sk_1,\sk_2,\sk_3)
& = 
	\int\, d^4x\, \epsilon^{ijk}\, \ScS(\sk_1|x)\,  \VV^{\eta_1}_i(\sk_2|x)\,  \partial_j\VV^{\eta_2}_k(\sk_3|x)	\\
\VVSb^{\eta_1\,\eta_2}(\sk_1,\sk_2,\sk_3)
& = 
	\int\, d^4x\, \epsilon^{ijk}\, \VV^{\eta_1}_i(\sk_1|x)\,  \VV^{\eta_2}_j(\sk_2|x)\, \partial_k\ScS(\sk_3|x)\\
&=
	- \VVSa^{\eta_2\eta_1}(\sk_3,\sk_2,\sk_1) + \VVSa^{\eta_1\eta_2}(\sk_3,\sk_1,\sk_2)
\end{split}
\end{equation}	
Finally, terms with two scalars contribute with two distinct structures, which are 
\begin{equation}\label{eq:VSSffactor}
\begin{split}
\VSSa^{\eta}(\sk_1,\sk_2,\sk_3)
& = 
	\int\, d^4x\, \epsilon^{ijk}\, \ScS(\sk_1|x)\,  \partial_i\, \ScS(\sk_2|x)\, \partial_j\VV^{\eta}_k(\sk_3|x)	\\	
\VSSb^{\eta}(\sk_1,\sk_2,\sk_3)
& = 
	\int\, d^4x\, \epsilon^{ijk}\, \VV^{\eta}_i(\sk_1|x)\,  \partial_j\ScS(\sk_2|x)\, \partial_k\ScS(\sk_3|x)	\\
&=
	- \VSSa^{\eta}(\sk_3,\sk_2,\sk_1)\,.
\end{split}
\end{equation}
One can simplify some of these structures using the identity~\eqref{eq:chiralcurl}.

\section{The mock tortoise coordinate}\label{sec:mockT}

We give here an analytic expression for  $\ctor(r)$, which is defined in~\eqref{eq:mockT}. As noted there, this is required to satisfy
\begin{equation}\label{eq:zetaBdy}
\lim_{r\rightarrow\infty+i0}\zeta(r) = 0\,, \qquad 
\lim_{r\rightarrow\infty-i0}\zeta(r) = 1\,, 
\end{equation}
The function $f$ has six simple zeros, and we need to orient the cuts from the integral appropriately. To achieve this, we factorize
\begin{equation}
\begin{split}
\frac{1}{r^2f(r)} = 
\frac{r^4}{(r^2-r_+^2)(r^2-r_{-}^2)(r^2+r_*^2)}\,, \\
\end{split}
\end{equation}
by setting 
\begin{equation}
r_{-} = \sqrt{\frac{-1+\sqrt{1+4Q^2}}{2}}\,r_+\,,\qquad 
r_{*} = \sqrt{\frac{1+\sqrt{1+4Q^2}}{2}}\,r_+\,,.
\end{equation}	
The integral can then be easily evaluated by partial fractions.

The branch cut from $r_+$, which is of physical interest, is oriented along the ray $r \in (r_+,\infty)$, while those from the other roots are directed away from this ray. This ensures that the along the keyhole contour depicted in~\cref{fig:mockt} only picks up the monodromy from the cut emanating from the outer horizon. A convenient parameterization turns out to be
\begin{equation}
\begin{split}
\ctor(r) 
&= 
	\frac{i}{4\pi}\left[4\,\tanh^{-1}\left(\frac{r}{r_+}\right)  
	-\frac{r_{*}}{2\,r_+}
	\left(\frac{1+Q^2-(1-Q^2)\sqrt{1+4\,Q^2}}{Q\sqrt{1+4\,Q^2}}\right)
	\left(\frac{i\pi}{2} + \tanh^{-1}\left(\frac{r_{-}}{r}\right)\right)\right. \\ 
&\qquad \qquad 
\left. 
	-\frac{r_{-}}{2\,r_+}\left(\frac{1+Q^2+(1-Q^2)\sqrt{1+4\,Q^2}}{Q\sqrt{1+4\,Q^2}}\right)\arctan\left(\frac{r}{r_*}\right)
\right] - \zeta_c\,,
\end{split}
\end{equation}
where $\zeta_c$  chosen to ensure $\ctor(\infty+i\epsilon)=0$.

%

\providecommand{\href}[2]{#2}\begingroup\raggedright\endgroup


\end{document}